\documentclass[smallcondensed,a4paper]{svjour3}

\usepackage[british]{babel}
\usepackage{amsmath}
\usepackage{listings}
\usepackage{lstcoq}
\usepackage{hyperref}
\usepackage{amssymb}
\usepackage{mathtools}
\usepackage{stmaryrd}
\usepackage{proof-dashed}
\usepackage[outline]{contour}
\usepackage{xspace}
\usepackage{xparse}
\usepackage{xpatch}
\usepackage[capitalize]{cleveref}
\usepackage{cite}
\usepackage[all]{xy}

\definecolor{dkviolet}{rgb}{.5,0,.5}
\definecolor{dkblue}{rgb}{0,0,.5}
\definecolor{dkgreen}{rgb}{0,.5,0}
\definecolor{dkred}{rgb}{.5,0,0}
\definecolor{ltblue}{rgb}{0,.4,.7}

\NewDocumentCommand{\epp}{m o o}
	{\llbracket{#1}\rrbracket%
		\IfNoValueF{#2}{_{#2}}%
		\IfNoValueF{#3}{^{#3}}}

\newcommand*{\rname}[1]{\ensuremath{\textsc{\MakeLowercase{#1}}}}
\newcommand*{\m}[1]{\mathsf{#1}}

\newcommand*{\pid}[1]{\ensuremath{\m{#1}}}
\newcommand*{\tto}{\ensuremath{\mathbin{\boldsymbol{\rightarrow}}}}
\newcommand*{\com}[2]{\ensuremath{#1 \tto #2}}
\newcommand*{\nil}{\ensuremath{\boldsymbol{0}}\xspace}
\newcommand*{\lto}[1]{\xrightarrow{#1}}
\newcommand*{\disjoint}{\mathbin{\#}}

\newcommand*{\psend}[1]{\ensuremath{{#1}!}}
\newcommand*{\precv}[1]{\ensuremath{{#1}?}}
\newcommand*{\proc}[2]{#1 [#2]}
\newcommand*{\pp}{\mathbin{\boldsymbol{|}}}

\newcommand*{\rulebreak}{\\[1em]}

\newcommand{\eoe}{\hfill $\triangleleft$}

\preto{\endexample}{\eoe}

\journalname{}

\begin{document}

\title{A Formal Theory of Choreographic Programming}
\author{Luís Cruz-Filipe \and Fabrizio Montesi \and Marco Peressotti}
\institute{Luís Cruz-Filipe \at Department of Mathematics and Computer Science\\
University of Southern Denmark\\
Campusvej 55\\
5230 Odense M\\
\email{lcf@imada.sdu.dk}
\and
Fabrizio Montesi \at Department of Mathematics and Computer Science\\
University of Southern Denmark\\
Campusvej 55\\
5230 Odense M\\
\email{fmontesi@imada.sdu.dk}
\and
Marco Peressotti \at Department of Mathematics and Computer Science\\
University of Southern Denmark\\
Campusvej 55\\
5230 Odense M\\
\email{peressotti@imada.sdu.dk}
}
\date{}
\maketitle

\begin{abstract}
Choreographic programming is a paradigm for writing coordination plans for distributed systems from a global point of view, from which correct-by-construction decentralised implementations can be generated automatically.

Theory of choreographies typically includes a number of complex results that are proved by structural induction.
The high number of cases and the subtle details in some of these proofs has led to important errors being found in published works.

In this work, we formalise the theory of a choreographic programming language in Coq.
Our development includes the basic properties of this language, a proof of its Turing completeness, a compilation procedure to a process language, and an operational characterisation of the correctness of this procedure.

Our formalisation experience illustrates the benefits of using a theorem prover: we get both an additional degree of confidence from the mechanised proof, and a significant simplification of the underlying theory.
Our results offer a foundation for the future formal development of choreographic languages.

\end{abstract}

\section{Introduction}
\label{sec:intro}
In the setting of concurrent and distributed systems, choreographic languages are used to define
interaction protocols that communicating processes should abide to~\cite{msc,bpmn,wscdl}.
These languages are akin to the ``Alice and Bob'' notation found in security protocols, and inherit
the key idea of making data communication manifest in programs~\cite{NS78}.
This is usually obtained through a linguistic primitive like \lstinline+Alice.e -> Bob.x+, read
``\lstinline+Alice+ communicates the result of evaluating expression \lstinline+e+ to
\lstinline+Bob+, which stores it in its local variable \lstinline+x+''.

In recent years, the communities of concurrency theory and programming languages have been prolific
in developing methodologies based on choreographies, yielding results in program verification,
monitoring, and program synthesis~\cite{Aetal16,Hetal16}.
For example, in \emph{multiparty session types}, types are choreographies used for checking
statically that a system of processes implements protocols correctly~\cite{HYC16}.
Further, in \emph{choreographic programming}, choreographic languages are elevated to full-fledged
programming languages~\cite{M13p}, which can express how data should be pre- and post-processed by
processes (encryption, validation, anonymisation, etc.).

Choreographic programming languages come with a procedure known as \emph{Endpoint Projection (EPP)}, which automatically synthesises executable code for each process described in a choreography, with
the guarantee that executing these processes together implements the communications prescribed in
the choreography~\cite{CHY12,CM13}.
These languages showed promise in a number of contexts, including parallel
algorithms~\cite{CM16}, cyber-physical systems~\cite{LNN16,LH17,GMP20}, self-adaptive
systems~\cite{DGGLM17}, system integration~\cite{GLR18}, information flow~\cite{LN15}, and the
implementation of security protocols~\cite{GMP20}.

EPP involves three elements: the source choreographic language, the target process language, and the compiler.
The interplay between these components, where a single instruction at the choreographic level might
be implemented by multiple instructions in the target language, makes the theory of choreographic
programming error-prone: for even simpler approaches, like abstract choreographies without
computation, it has been recently discovered that a few key results published in peer-reviewed
articles do not hold and their theories required adjustments~\cite{SY19}, raising concerns about the
soundness of these methods.

This article presents a formalisation of a core theory 
of choreographic programming in the theorem prover Coq, the process of developing this formalisation, the challenges encountered, and how tackling these challenges led to improvements of the original theory.

\paragraph{A note on the process.}
We argue that computer-aided verification can be successfully applied to the study of choreographies and to provide solid
foundations for future developments.
To substantiate this claim, we summarise the story behind this article, which illustrates how interactive theorem proving can do more than just checking what we already know.

Our starting point was the theory of Core Choreographies (CC), a minimalistic language that the first two authors previously proposed for the study of choreographic programming~\cite{CM20}.
CC includes only the essential features of choreographic languages and minimal
computational capabilities at processes (computing the successor of a natural number and deciding
equality of two natural numbers), yet it is expressive enough to be Turing complete.

We started formalising CC in Coq in late 2018.
In mid-2019, we gave an informal progress report on the promising status of the formalisation at the TYPES conference~\cite{CMP19}.
Unfortunately, we soon stumbled upon an unexpected source of complexity for the formalisation: a set of
term-rewriting rules for a precongruence relation used in the semantics of the language for
(i)~expanding procedure calls and (ii)~reshuffling
independent communications to model concurrent execution.
In addition to being time consuming, reasoning with precongruence systematically made the formalisation
significantly more complicated than the development in~\cite{CM20} (for a more technical
discussion, see~\cref{sec:chor-discussion}).

At the time, the second author was responsible for a Master course on theory of choreography for students in Computer Science.
It quickly became apparent that the technical aspects (including, but not only, structural precongruence) that complicated the formalisation of CC were also the most challenging for the students.
This observation led that author to develop an alternative theory of CC for his course material that
dispenses with these problematic notions without changing its essence~\cite{M22}.
The formalisation in this article uses this revised choreography theory.

Thus, our work also shows that theorem proving can be used
in research: the insights obtained while doing this formalisation led to changes in the original theory.
We show that this did not come at the cost of expressive power: the original proof of Turing completeness from~\cite{CM20} still works for the theory in~\cite{M22} without essential changes~\cite{CMP21a}.
Furthermore, formalising the theory also allowed us to identify unnecessary assumptions in some
lemmas, yielding stronger results.

\paragraph{Publication history and contribution.}
As mentioned previously, a first informal progress report on this formalisation was presented at the
TYPES conference in 2019~\cite{CMP19}, following an approach that later turned to be unfeasible.
The first formalisation of the choreographic language, including the proof of Turing completeness,
was presented in~\cite{CMP21a}, while the formalisation of EPP appeared
originally in~\cite{CMP21b}.
The current presentation discusses an updated formalisation, which (i)~no longer uses Coq's module
system
and (ii)~differs significantly in the treatment of partial functions, which significantly
simplifies the definition of EPP.
We do not discuss the formalisation of the proof of Turing completeness, as this is essentially
unchanged from~\cite{CMP21a}.
Instead, we place a stronger emphasis on the formalisation challenges compared to the works cited.

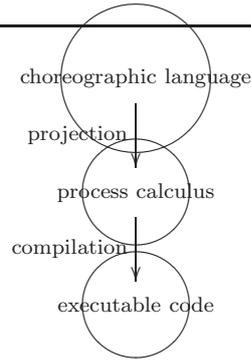
\begin{figure}
  \[\xymatrix{
    *++[o][F-:3pt]\txt{choreographic language}
    \ar[d]_{\txt{projection}}
    \\
    *++[o][F-:3pt]\txt{process calculus}
    \ar[d]_{\txt{compilation}}
    \\
    *++[o][F-:3pt]\txt{executable code}
  }\]

  \caption{Two-stage compilation process from choreographies to executable code.}
  \label{fig:missing}
\end{figure}

\paragraph{The big picture.}
This work is the first step towards a more ambitious goal: the development of a certified framework for choreographic programming.
At a later stage, we plan on developing compilers that can translate the process implementations generated by EPP into executable code in different programming languages (see \cref{fig:missing}).
This would yield end-to-end compilation from choreographies to actual executable code.

Our goal motivated two important design choices in the current work that are not present in~\cite{CMP21a,CMP21b}.
First, we want to extract a correct implementation of EPP from our formalisation; this required moving away from the usage of Coq's module system, as it does not support extraction.
Second, we introduce the possibility of annotating terms in choreographies with data that may be needed for (second-stage) compilation to executable programming languages.

\paragraph{Structure.}
A full understanding of the more technical details of our formalisation benefits from some background knowledge on choreographies.
For convenience, \Cref{sec:background} features a short introduction to the main intuitions and
results of choreography theory, which can be skipped by readers familiar with the topic.
Our choreographic language (syntax and semantics) is presented together with its Coq formalisation
in \Cref{sec:cc}, where it is also shown that it enjoys the usual properties of choreographic languages.
\Cref{sec:sp} defines the target process language, together with its semantics.
EPP is formalised in \Cref{sec:epp}, and its soundness and completeness are
discussed in \Cref{sec:epp-thm}.
We review related formalisation efforts in
\Cref{sec:rw}, before concluding in \Cref{sec:concl}.

The source code of our Coq development is available at \cite{CMP22-source}.

\section{Background: Choreographic Languages and Endpoint Projection}
\label{sec:background}
In this section we describe the language of Simple Choreographies~\cite{M22}, which introduces the basic principles of choreographies and EPP. We include this material to make our development accessible to the reader not familiar with the topic, but it is not directly used in our development.

\subsection{Simple Choreographies}
Simple Choreographies can express finite sequences of communications between processes.
Processes are identified by names (\pid p, \pid q, etc.).
Choreographies, ranged over by $C$, are constructed according to the following grammar.
\begin{align*}
C & \Coloneqq \com{\pid p}{\pid q};C \mid \nil
\end{align*}

A choreography $\com{\pid p}{\pid q};C$ represents a communication from a process \pid p to a process \pid q with continuation $C$; $\nil$ is the terminated choreography. We omit trailing $\nil$s in examples.

\begin{example}[Ring protocol~\cite{M22}]
The choreography below describes a ring protocol among three participants: \pid{Alice} communicates to \pid{Bob}; then \pid{Bob} communicates to \pid{Carol}; and finally \pid{Carol} communicates back to \pid{Alice}.
\begin{equation}\label{eq:ring-3}
\com{\pid{Alice}}{\pid{Bob}}; \com{\pid{Bob}}{\pid{Carol}};
\com{\pid{Carol}}{\pid{Alice}}
\end{equation}
\end{example}

The semantics of Simple Choreographies is given as the labelled transition system induced by the rules displayed in \Cref{fig:sc-semantics}. Transition labels have the form \com{\pid p}{\pid q}, allowing for observing the communications performed by a choreography.
\begin{figure}
\begin{gather*}
\infer[\rname{Com}]{
	\com{\pid p}{\pid q}; C \lto{\com{\pid p}{\pid q}} C
}{}
\quad
\infer[\rname{Delay}]{
	\com{\pid p}{\pid q}; C \lto{\mu} \com{\pid p}{\pid q}; C'
}{
	C \lto{\com{\pid r}{\pid s}} C'
	&
	\{\pid p, \pid q\} \disjoint \{\pid r, \pid s\}
}
\end{gather*}
\caption{Semantics of Simple Choreographies.}
\label{fig:sc-semantics}
\end{figure}

Rule \rname{Com} models the execution of a communication at the beginning of a choreography.
Rule \rname{Delay}, instead, allows for performing a transition within the continuation of a choreography, provided that the transition does not involve any of the processes in preceding instructions.
This rule captures the fact that processes run independently of each other, and thus choreographic instructions can be executed out-of-order. The independence requirement is captured by the the side-condition $\{\pid p, \pid q\} \disjoint \{\pid r, \pid s\}$, where $\disjoint$ relates disjoint sets.

\begin{example}[Ring protocol, continued~\cite{M22}]
Let $C$ be the choreography in~\eqref{eq:ring-3}.
Then, by rule \rname{Com}, we have the following chain of transitions.
\[
C \lto{\com{\pid{Alice}}{\pid{Bob}}}
\com{\pid{Bob}}{\pid{Carol}};
\com{\pid{Carol}}{\pid{Alice}}
\lto{\com{\pid{Bob}}{\pid{Carol}}}
\com{\pid{Carol}}{\pid{Alice}}
\lto{\com{\pid{Carol}}{\pid{Alice}}}
\nil
\]
These communications cannot be executed out-of-order, because of the chain of causality between them: each instruction involves a process that needs to participate in a previous instruction.
\end{example}

\begin{example}\label{ex:oo-2}
Consider now the choreography (inspired from the factory examples in \cite{M22}), which models a system where two ``ordering'' processes $\pid o_1$ and $\pid o_2$ independently communicate two respective orders to the servers $\pid s_1$ and $\pid s_2$.
\begin{equation}\label{eq:oo-2}
\com{\pid o_1}{\pid s_1}; \com{\pid o_2}{\pid s_2}
\end{equation}
The following derivation shows that $\com{\pid o_2}{\pid s_2}$ can be executed first.
\[
\infer[\rname{Delay}] {
  \com{\pid o_1}{\pid s_1}; \com{\pid o_2}{\pid s_2}
  \lto{\com{\pid o_2}{\pid s_2}}
  \com{\pid o_1}{\pid s_1}
}{
  \infer[\rname{Com}] {
    \com{\pid o_2}{\pid s_2}
    \lto{\com{\pid o_2}{\pid s_2}}
    \nil
  }{}
  &
  \{ \pid o_1, \pid s_1 \} \disjoint \{\pid o_2, \pid s_2\}
}
\]
\end{example}

\subsection{Simple Processes}
Implementations of Simple Choreographies are modelled in a process language called Simple Processes~\cite{M22}.
First, we define a grammar for writing process behaviours.
\begin{align*}
P,Q,R & \Coloneqq \psend{\pid p};P \mid \precv{\pid p};P \mid \nil
\end{align*}
These actions are the local counterparts to the communication action in choreographies.
A send action $\psend{\pid p}$ sends a message to a process $\pid p$, and the dual receive action $\precv{\pid p}$ receives a message from a process $\pid p$. The term $\nil$ is the terminated process.

Processes are composed into networks ($N$, $M$, etc.), which are maps from process names to processes.
We introduce some notation: $\nil$ is the terminated network, where all process names are mapped to $\nil$; $\proc{\pid p}{P}$ is the network where $\pid p$ is mapped to $P$ and all other process names are mapped to $\nil$; and $N \pp M$ (``$N$ parallel $M$'') is the union of $N$ and $M$, assuming that their supports\footnote{The support of a network is the set of all processes not mapped to $\nil$.} are disjoint.
Under extensional equality of functions, the set of networks equipped with parallel composition forms a partial commutative monoid with $\nil$ as identity element: $N \pp \nil = N$, $N \pp M = M \pp N$, and $N_1 \pp (N_2 \pp N_3) = (N_1 \pp N_2) \pp N_3$~\cite{M22}.

\begin{example}
The following network implements the choreography in~\eqref{eq:ring-3}.
\begin{equation}\label{eq:ring-3-network}
\proc{\pid{Alice}}{
  \psend{\pid{Bob}};
  \precv{\pid{Carol}}
} \pp
\proc{\pid{Bob}}{
  \precv{\pid{Alice}};
  \psend{\pid{Carol}}
} \pp
\proc{\pid{Carol}}{
  \precv{\pid{Bob}};
  \psend{\pid{Alice}}
}
\end{equation}
\end{example}

\begin{figure}
\begin{gather*}
\infer[\rname{Com}]{
	\proc{\pid p}{\psend{\pid q}; P}
	\pp
	\proc{\pid q}{\precv{\pid p}; Q}
	\lto{\com{\pid p}{\pid q}}
	\proc{\pid p}{P} \pp \proc{\pid q}{Q}
}{}
\quad
\infer[\rname{Par}]{
	N \pp M \lto{\com{\pid p}{\pid q}} N' \pp M
}{
	N \lto{\com{\pid p}{\pid q}} N'
}
\end{gather*}
\caption{Semantics of Simple Processes.}
\label{fig:sp-semantics}
\end{figure}
The semantics of Simple Processes is given by the transition rules in \Cref{fig:sp-semantics}. Rule \rname{Com} synchronises processes with matching send and receive actions. Rule \rname{Par} allows for parallel execution.

\begin{example}
The transitions of the choreography in~\eqref{eq:ring-3} coincide with those of the network in~\eqref{eq:ring-3-network}. Technically, the labelled transition systems generated by the choreography and the network are isomorphic, showing that the network is indeed a precise implementation of the choreography.
\end{example}

\begin{example}
Out-of-order execution for choreographies corresponds to parallelism at the level of networks.
The following network implements the choreography in~\eqref{eq:oo-2}.
\begin{equation}\label{eq:oo-2-network}
\proc{\pid o_1}{\psend{\pid s_1}}
\pp
\proc{\pid o_2}{\psend{\pid s_2}}
\pp
\proc{\pid s_1}{\precv{\pid o_1}}
\pp
\proc{\pid s_2}{\precv{\pid o_2}}
\end{equation}
Using rule \rname{Par} and the monoidal structure of parallel composition, the network can start by executing either the communication between $\pid o_1$ and $\pid s_1$ or the one between $\pid o_2$ and $\pid s_2$.
\end{example}

\subsection{Endpoint projection}
\label{sec:bg-epp}

In general, writing correct implementations of protocols is hard, especially for more expressive choreographic languages as the one that we use later in this article.
Endpoint projection (EPP) is a mechanical procedure for translating choreographies into networks by splitting choreographic terms into their local counterparts~\cite{CHY12,CM13,HYC16,CM20,M22}.
The idea is that given a choreography $C$ and a process $\pid p$, we first compute the process term $\epp{C}[\pid p]$ that implements the actions that \pid p should perform to implement its part in $C$. Then, EPP is defined as the parallel composition of all such terms.

\begin{figure}
\begin{align*}
\epp{\com{\pid p}{\pid q}; C}[\pid r] & =
	\begin{cases}
		\psend{\pid q};\epp{C}[\pid r] & \text{if } \pid r = \pid p\\
		\precv{\pid p};\epp{C}[\pid r] & \text{if } \pid r = \pid q\\
		\epp{C}[\pid r] & \text{otherwise}
	\end{cases}
&
\epp{\nil}[\pid r] & = \nil
\end{align*}
\caption{Process projection for Simple Choreographies.}
\label{fig:p-proj}
\end{figure}
In the case of Simple Choreographies and Simple Processes, the process projection map $\epp{C}[\pid p]$ is defined in a natural way by the recursive equations in \Cref{fig:p-proj}.
In particular, a communication term $\com{\pid p}{\pid q}; C$ is projected to a send action and the projection of the continuation if we are projecting the sender (first case), a receive action and the projection of the continuation if we are projecting the receiver, or just the projection of the continuation if we are projecting a process that is not involved in the communication. 

Given a choreography $C$, its EPP $\epp C$ is defined as the network $\epp{C}(\pid p) = \epp{C}[\pid p]$.
This network is a correct implementation of $C$.
\begin{theorem}[Correctness of EPP~\cite{M22}]
\label{thm:epp-sc}
The following statements hold for every choreography $C$ and transition label $\mu$ in the language of Simple Choreographies.
\begin{description}
\item[\em\bf Completeness] For any $C'$, if $C \lto\mu C'$ then $\epp{C} \lto\mu \epp{C'}$.
\item[\em\bf Soundness] For any $N$, if $\epp{C} \lto\mu N$ then $C \lto\mu C'$ for some $C'$ such that $N = \epp{C'}$.
\end{description}
\end{theorem}

\begin{example}
The networks in~\eqref{eq:ring-3-network} and~\eqref{eq:oo-2-network} are, respectively, the EPPs of the choreographies in~\eqref{eq:ring-3} and~\eqref{eq:oo-2}.
\end{example}

\subsection{Taking Stock}
\Cref{thm:epp-sc} is used to prove other notable results given by the choreographic approach, such as \emph{deadlock-freedom}.
A deadlocked network is one that is not terminated but cannot make any transitions, typically because all processes are waiting for someone else.
Even in a simplistic process language such as Simple Processes, we can write deadlocked networks, such as:
\[
\proc{\pid p}{\precv{\pid q}} \pp \proc{\pid q}{\precv{\pid p}}.
\]
Here, \pid p and \pid q are both waiting for each other, and therefore the network will never be able to proceed.

Since communication terms in choreographies specify simultaneously what sender and receiver processes are involved, choreographies cannot describe deadlocks, a property known as \emph{deadlock-freedom by design}~\cite{CM13}. As a consequence of \cref{thm:epp-sc}, the networks generated by EPP can never become deadlocked.

The choreographic language that we consider in the rest of our article is more expressive than Simple Choreographies, as it includes features that are important for modelling realistic protocols. However, the general structure of the development follows the roadmap given in this section, albeit with a much higher level of complexity.

\section{Core Choreographies}
\label{sec:cc}

We introduce Core Choreographies (CC), the choreographic language that we work with, and its formalisation.
At the end of this section, we discuss how the formalisation process guided the evolution of the language from its original presentation in~\cite{CM20} to its present form, which is closer to the style of~\cite{M22}.

In CC, processes can perform point-to-point communications and have storage.
Communicated messages can be either values, which are computed by evaluating local expressions, or labels (tags, or constants) from a fixed set $\{\coc{left}, \coc{right}\}$.\footnote{Restricting the set of labels to two elements is standard practice~\cite{CM20,CP10}.}
Additionally, choreographies can include conditionals based on Boolean expressions and invoke recursive procedures.

\subsection{Preliminaries}

Choreographies are parameterised on a signature, which defines the types for process names (processes for short) \lstinline+pid+, local variables \lstinline+var+ (used to access the processes' storage), values \lstinline+val+, expressions \lstinline+expr+, Boolean expressions \lstinline+bexpr+, and procedure names \lstinline+recvar+ (from \emph{recursion variables}).
Signatures also include types for (user-defined) annotations \lstinline+ann+ (as discussed in \cref{sec:intro}). Since the types of expressions and values are parameters, signatures also need to specify the evaluation functions mapping expressions to values and Boolean expressions to Booleans.
We fix a signature \lstinline+Sig+ and introduce abbreviations \lstinline+Pid := (pid Sig)+ and similarly for all other parameters for convenience.

The first seven parameters are datatypes equipped with a decidable equality.
Since we are targetting extraction, which is not compatible with modules, we reimplemented \coc{DecType} as a record type consisting of exactly these two components, and reproved the lemmas about decidable equality from the Coq standard library.
We also show that the Cartesian product of two \lstinline+DecType+s can be made into a \lstinline+DecType+, and we define a two-element decidable type \lstinline+Label+ whose elements are the two labels \coc{left} and \coc{right}.

Evaluation functions are again records.
The first element is a function that takes an expression and a mapping from a process's variables to values, and returns a value (possibly of a different type as the one stored locally).
The second element is a proof that the value returned by evaluation does not change if the mapping from variables to values is replaced by an extensionally equivalent one.

The type \lstinline+State := Pid -> Var -> Value+ models the memory state of the set of all processes.\footnote{In the formalisation, the type \lstinline+Var -> Value+ is given the name \lstinline+LState+ (for ``local state''), but since local states are unused elsewhere we do not discuss them here.}
We define extensional equality on states, written \coc{[==]}, and prove that it is an equivalence relation.
Furthermore, we define an operation \lstinline+s[[p,x => v]]+ for updating the state \lstinline+s+ with the assignment of value \lstinline+v+ to process \lstinline+p+'s variable \lstinline+x+, and prove a number of useful rewriting lemmas.

\subsection{Syntax}
\label{sec:chor-syntax}

Choreographies are defined inductively by the following grammar.\footnote{Throughout this article, we use the pretty-printing rules defined in the Coq formalisation so that the correspondence between the informal mathematical presentation and the formal results is clear.}

\begin{lstlisting}
  $\eta$ $\Coloneqq$ p#e --> qdollarx $\mid$ p --> q[l]
  C $\Coloneqq$ $\eta$@a;; C $\mid$ If p ?? b Then C1 Else C2 $\mid$ Call X $\mid$ RT_Call X ps C $\mid$ End
\end{lstlisting}

Here, \lstinline+p,q:Pid+ are processes, \lstinline+e:Expr+ is an expression, \lstinline+x:Var+ is a variable, \lstinline+l:Label+ is a label, \lstinline+a:Ann+ is an annotation, \lstinline+b:BExpr+ is a Boolean expression, \lstinline+X:RecVar+ is a procedure name, and \lstinline+ps:list Pid+ is a list of processes.

The terms denoted $\eta$ are called \emph{interactions}; for many results, it is convenient that they form their own type.
Term \lstinline+p#e --> qdollarx+ is a value communication, where \coc{p} communicates the result of evaluating \coc{e} to \coc{q}, which stores it in its local variable \coc{x}.
Term \lstinline+p --> q[l]+ is a label selection, where \coc{p} communicates label \coc{l} to \coc{q}.

Label selections are used in conjunction with conditionals.
In a conditional \coc{If p ?? b Then C1 Else C2}, the evolution of the choreography is determined by the outcome of evaluating the Boolean expression \coc{b} at \coc{p}.
Other processes that need to know which branch was chosen (\emph{knowledge of choice}~\cite{CDP11}) can get this information through the reception of label \coc{left} or \coc{right} from \coc{p}.

Interactions are paired with annotations (\coc{a}), which are not used in this work.
They are meant to include additional information that may be needed in subsequent processing steps, such as documentation or the second-stage compilation mentioned in \cref{sec:intro}.
We omit annotations in all our examples.

Term \lstinline+Call X+ invokes the procedure named \lstinline+X+.
A procedure may involve several processes, and the semantics of CC allows each process to join the procedure only when needed.
The \emph{runtime} term \lstinline+RT_Call X ps C+ represents this intermediate situation: execution of procedure \coc{X} has already evolved to \coc{C}, but the processes in \coc{ps} have not yet joined it.
Runtime terms are not meant to be written by programmers: they are auxiliary terms generated by the semantics.

The grammar of choreographies is defined as the following inductive types.
\begin{lstlisting}
Inductive Eta : Type :=
| Com : Pid -> Expr -> Pid -> Var -> Eta
| Sel : Pid -> Pid -> Label -> Eta.

Inductive Choreography : Type :=
| Interaction : Eta -> Ann -> Choreography -> Choreography
| Cond : Pid -> BExpr -> Choreography -> Choreography -> Choreography
| Call : RecVar -> Choreography
| RT_Call : RecVar -> (list Pid) -> Choreography -> Choreography
| End : Choreography.
\end{lstlisting}

A set of procedure definitions, formalised as type \lstinline+DefSet+, is a mapping assigning to each \lstinline+RecVar+ a list of processes and a choreography; intuitively, the list contains the processes that are used in the procedure.
A \lstinline+Program+ is a pair containing a set of procedure definitions and the choreography to be executed at the start, also called the \emph{main} choreography.

\begin{lstlisting}
Definition DefSet := RecVar -> (list Pid)*Choreography.
Definition Program := DefSet * Choreography.
\end{lstlisting}

We write \lstinline+Procedures P+ and \lstinline+Main P+ for, respectively, the set of procedure definitions and the main choreography in a program \lstinline+P+ (so \lstinline+Procedures+ and \lstinline+Main+ are simply aliases for the corresponding projections).
Likewise, \lstinline+Vars P X+ and \lstinline+Procs P X+ denote the list of processes and the definition of a particular procedure \lstinline+X+ within \lstinline+P+.
Finally, \lstinline+Names D+ is the function mapping each variable \lstinline+X+ to the set of processes that it uses according to \coc{D:DefSet}.

\begin{example}[Distributed Authentication]
  \label{ex:authentication_chor}
  The choreography \lstinline+C1+ below describes a multiparty authentication scenario where an identity provider \lstinline+ip+ authenticates a client \lstinline+c+ to server \lstinline+s+.
  (For convenience, we name some of the subterms in the choreography.)
  \begin{lstlisting}
 C1 := c#credentials --> ipdollarx;; If ip ?? (check x) Then C1t Else C1e
C1t := ip --> s[left];; ip --> c[left];; s#token --> cdollart;; End
C1e := ip --> s[right];; ip --> c[right];; End
  \end{lstlisting}
  \lstinline+C1+ starts with \lstinline+c+ communicating its 
  \coc{credentials} to \lstinline+ip+, which stores them in \lstinline+x+.
  Then, \lstinline+ip+ checks whether the received credentials are valid by evaluating the Boolean expression \lstinline+check x+, and signals the result to \lstinline+s+ and \lstinline+c+ by selecting \lstinline+left+ when the credentials are valid (\lstinline+C1t+) and \lstinline+right+ otherwise (\lstinline+C1e+).
  In the first case, the server communicates a \coc{token} to \coc{c}, otherwise the choreography simply terminates.

  The selections from \coc{ip} to \coc{s} and \coc{c} address knowledge of choice, as previously described.
\end{example}

\paragraph{Well-formedness.}
There are a number of well-formedness requirements on choreographies, which can be grouped in three categories.
\begin{enumerate}
\item Intended use of choreographies.
  Interactions must have distinct processes (there are no self-communication), e.g., \lstinline+p#e --> pdollarx+ is disallowed.

\item Intended use of runtime terms.
  Procedure definitions may not contain runtime terms.
  \lstinline+Main P+ may include subterms \lstinline+RT_Call X ps C+, but \lstinline+ps+ must be nonempty and include only process names that occur in \lstinline+Vars P X+.

\item Design choices in the formalisation.
  The processes in \lstinline+Vars X+ include all processes that are used in \lstinline+Procs X+.
\end{enumerate}
Well-formedness is essential in the proof of correctness of EPP (\Cref{sec:epp-thm}).

We start by formalising the different properties of choreographies separately:
\begin{itemize}
\item\lstinline+initial C+ holds if \lstinline+C+ does not contain runtime terms (\coc{RT_Call});
\item\lstinline+no_self_comm C+ holds if \lstinline+C+ contains no self-communications;
\item\lstinline+no_empty_ann C+ holds if all runtime terms in \coc{C} have nonempty lists of process names.
\end{itemize}
These properties are defined recursively over \lstinline+C+ in the natural way.
Well-formedness of choreographies \lstinline+Choreography_WF+ is defined as the conjunction of the last two properties.

Well-formedness of programs also takes into account the additional requirements on the lists of processes annotating runtime terms.
Specifically, in a program \lstinline+P+, the choreography \lstinline+Main P+ must be consistently annotated with respect to \lstinline+Vars P+: in any subterm \lstinline+RT_Call X ps C'+ in \lstinline+Main P+, the list \lstinline+ps+ only contains processes appearing in \lstinline+Vars P X+.
This property is written as \lstinline+consistent (Vars P) (Main P)+, where predicate
\begin{lstlisting}
consistent: (RecVar -> list Pid) -> Choreography -> Prop
\end{lstlisting}
is defined inductively in the expected way.
Also, the set of procedure definitions in \lstinline+P+ must be well-annotated: if \lstinline+Procedures P X=(ps,C)+, then the set of processes used in \lstinline+C+ must be a nonempty subset of \coc{ps}.\footnote{We defined suggestive notations \lstinline+[C]+, \lstinline+[U]+, \lstinline+[$\setminus$]+ and \lstinline+[#]+ for the set operations we use.}
\begin{lstlisting}
Definition well_ann (P:Program) (X:RecVar) : Prop :=
  Vars P X <> nil /\ CCC_pn (Procs P X) (Vars P) [C] Vars P X.
\end{lstlisting}

The last definition uses function \lstinline+CCC_pn+, which computes the set of processes occurring in a choreography, given the set of processes used in each procedure.
It generalises to \coc{CCP_pn}, which computes the set of processes occurring in a well-annotated program.

Using these ingredients, we define well-formedness of programs as follows.
\begin{lstlisting}
Definition Program_WF (P:Program) : Prop :=
  Choreography_WF (Main P) /\ consistent (Vars P) (Main P) /\
  forall X, no_self_comm (Procs P X) /\ initial (Procs P X) /\ well_ann P X.
\end{lstlisting}
Since \coc{initial} choreographies do not include runtime terms, this definition also implies that all procedure definitions are well-formed.

\begin{example}
  \label{ex:file_transfer_chor}
  Let \lstinline!Defs:DefSet! map \lstinline!FileTransfer! to the pair consisting of the process list \lstinline!c :: s :: nil! and the following choreography.
  \begin{lstlisting}
s.(file, check) --> c.x;;                     (* send file and check data          *)
If c.(crc(fst(x)) == snd(x))                (* cyclic redundancy check           *)
  Then c --> s[left];; End                    (* file received correctly, end      *)
  Else c --> s[right];; Call FileTransfer     (* errors detected, retry            *)
\end{lstlisting}
  \lstinline+FileTransfer+ describes a file transfer protocol between a server \lstinline+s+ and a client \lstinline+c+ using Cyclic Redundancy Checks (\lstinline+crc+) to detect errors from a noisy channel.

  Assuming that \lstinline+Defs+ maps all other procedure definitions to \lstinline+End+, the program \lstinline+P=(Defs,Call FileTransfer)+ satisfies \lstinline+Program_WF P+.
\end{example}

Recall that our long-term future goal is to apply program extraction to this formalisation, and then use the result in tools.
Many of the results that we show later only hold for well-formed programs, and any tool built on our theory should be able to validate that its input is well-formed.
However, due to the quantification over all procedure names, well-formedness of programs is in general not decidable.
In practice, though, choreographic programs only use a finite number of procedures; if these are known, well-formedness becomes decidable.

This observation motivates the definition of a recursive predicate
\begin{lstlisting}
used_procedures_C : Choreography -> list RecVar -> Prop
\end{lstlisting}
such that \coc{used_procedures_C C Xs} holds iff \coc{C} only calls procedures in \coc{Xs} (directly).
This is generalised to programs by requiring that all procedures in \coc{Xs} also satisfy the same property, and additionally that all procedures not in \coc{Xs} be defined as \coc{End}.
\begin{lstlisting}
Definition used_procedures (P:Program) (Xs:list RecVar) :=
  used_procedures_C (Main P) Xs /\
  forall X, (In X Xs -> used_procedures_C (Procs P X) Xs)
    /\ (~In X Xs -> Procs P X = End /\ Vars P X <> nil).
\end{lstlisting}
The requirement \coc{Vars P X <> nil} for procedures not in \coc{Xs} is included to ensure well-formedness.
From this, we can prove decidability of well-formedness.
\begin{lstlisting}
Lemma Program_WF_dec : forall P Xs, used_procedures P Xs ->
  {Program_WF P} + {~Program_WF P}.
\end{lstlisting}
Applying this lemma in extracted code requires knowing a suitable set \coc{Xs}.
While we cannot automatically verify that this set satisfies \coc{used_procedures P Xs}, it is very reasonable to trust that a correct one has been provided: typically, the relevant procedures used in a program are written down explicitly, making it straightforward to list them.

An alternative approach would be requiring the set of procedure names to be finite.
This is closer in spirit to the pen-and-paper presentations of choreographic languages -- even if procedure names are taken from an infinite set, only a finite number of them can be used in a concrete program~\cite{CM20}.
We chose the present approach for simplicity, as working with finite sets in Coq is notoriously cumbersome.

\subsection{Semantics}
\label{sec:chor-sem}

The semantics of CC is defined by means of labelled transition systems, in three layers.
At the lowest layer, we define the transitions that a choreography can make (\lstinline+CCC_To+), parameterised on a set of procedure definitions; then we pack these transitions into the more usual presentation -- as a labelled relation \lstinline+CCP_To+ on \emph{configurations} (pairs program/state).
Finally, we define multi-step transitions \lstinline+CCP_ToStar+ as the transitive and reflexive closure of the transition relation.
This layered approach makes proofs about transitions cleaner, allowing us to separate the different levels of induction.

\paragraph{Transition labels.}
Each layer of the semantics has its type of transition labels.
For the lower level, we define an inductive type \lstinline+RichLabel+ whose constructors reflect the possible actions a choreography can take: value communications, label selections, reducing a conditional, or locally joining a procedure call.

The second layer uses the type \lstinline+TransitionLabel+ of labels corresponding to the observable actions.
The two types are connected by a function \lstinline+forget:RichLabel -> TransitionLabel+.
Labels in the third layer are simply lists of \lstinline+TransitionLabel+s.
\begin{lstlisting}
Inductive RichLabel : Type :=
| RL_Com (p:Pid) (v:Value) (q:Pid) (x:Var) : RichLabel
| RL_Sel (p:Pid) (q:Pid) (l:Label) : RichLabel
| RL_Cond (p:Pid) : RichLabel
| RL_Call (X:RecVar) (p:Pid) : RichLabel.

Inductive TransitionLabel : Type :=
| TL_Com (p:Pid) (v:Value) (q:Pid) : TransitionLabel
| TL_Sel (p:Pid) (q:Pid) (l:Label) : TransitionLabel
| TL_Tau (p:Pid) : TransitionLabel.
\end{lstlisting}

Pen-and-paper presentations only include \coc{TransitionLabel}s, which capture what can be observed in transitions without revealing syntactic information about the choreography.
However, in Coq, this information is needed to obtain induction hypotheses that are strong enough for our development, which is why we have introduced \coc{RichLabel}s.

The transition relations are defined inductively by the rules in~\Cref{fig:CCC_To,fig:CCP_To,fig:CCP_ToStar}.
For readability, we present them in a more standard rule notation -- below, we exemplify how they correspond to constructors in the formalisation.
We also introduce suggestive notations for all these relations: \lstinline+<<C,s>> --[rl,D]--> <<C',s'>>+ stands for \lstinline+(CCC_To D C s rl C' s')+ (this relation is parameterised on \lstinline+D:DefSet+ for dealing with procedure calls); \lstinline"c --[tl]--> c'" stands for \lstinline+(CCP_To c tl c')+, where \lstinline+c,c':Configuration+ are pairs containing a \lstinline+Program+ and a \lstinline+State+; and \lstinline"c --[ts]-->** c'" stands for \lstinline+(CCP_ToStar c ts c')+.

\begin{figure}
  \begin{gather*}
    \infer[\mbox{\lstinline+C_Com+}]
          {\mbox{\lstinline+<<p#e --> qdollarx@a;; C,s>> --[RL_Com p v q x, D]--> <<C,s'>>+}}
          {\mbox{\lstinline+v := eval_on_state Ev e s p+}
            & \mbox{\lstinline+s' [==] s[[q,x => v]]+}}
    \rulebreak
    \infer[\mbox{\lstinline+C_Sel+}]
          {\mbox{\lstinline+<<p --> q[l]@a;; C,s>> --[RL_Sel p q l, D]--> <<C,s'>>+}}
          {\mbox{\lstinline+s [==] s'+}}
    \rulebreak
    \infer[\mbox{\lstinline+C_Then+}]
          {\mbox{\lstinline+<<If p??b Then C1 Else C2,s>> --[RL_Cond p, D]--> <<C1,s'>>+}}
          {\mbox{\lstinline+eval_on_state BEv b s p = true+}
            & \mbox{\lstinline+s [==] s'+}}
    \rulebreak
    \infer[\mbox{\lstinline+C_Else+}]
          {\mbox{\lstinline+<<If p??b Then C1 Else C2,s>> --[RL_Cond p, D]--> <<C2,s'>>+}}
          {\mbox{\lstinline+eval_on_state BEv b s p = false+}
            & \mbox{\lstinline+s [==] s'+}}
    \rulebreak
    \infer[\mbox{\lstinline+C_Delay_Eta+}]
          {\mbox{\lstinline+<<eta@ann;; C,s>> --[t,D]--> <<eta@ann;; C',s'>>+}}
          {\mbox{\lstinline+disjoint_eta_rl eta t+}
            & \mbox{\lstinline+<<C,s>> --[t,D]--> <<C',s'>>+}}
    \rulebreak
    \infer[\mbox{\lstinline+C_Delay_Cond+}]
          {\mbox{\lstinline+<<If p??b Then C1 Else C2,s>> --[t,D]--> <<If p??b Then C1' Else C2',s'>>+}}
          {\mbox{\lstinline+disjoint_p_rl p t+}
            & \begin{array}c
                \mbox{\lstinline+<<C1,s>> --[t,D]--> <<C1',s'>>+} \\
                \mbox{\lstinline+<<C2,s>> --[t,D]--> <<C2',s'>>+}
          \end{array}}
    \rulebreak
    \infer[\mbox{\lstinline+C_Delay_Call+}]
          {\mbox{\lstinline+<<RT_Call X ps C,s>> --[t,D]--> <<RT_Call X ps C',s'>>+}}
          {\mbox{\lstinline+disjoint_ps_rl ps t+}
            & \mbox{\lstinline+<<C,s>> --[t,D]--> <<C',s'>>+}}
    \rulebreak
    \infer[\mbox{\lstinline+C_Call_Local+}]
          {\mbox{\lstinline+<<Call X,s>> --[RL_Call X p,D]--> <<snd (D X),s'>>+}}
          {\mbox{\lstinline+s [==] s'+}
            & \mbox{\lstinline+[#](fst (D X)) = 1+}
            & \mbox{\lstinline+In p (fst (D X))+}}
    \rulebreak
    \infer[\mbox{\lstinline+C_Call_Start+}]
          {\mbox{\lstinline+<<Call X,s>> --[RL_Call X p,D]--> <<RT_Call X (fst (D X)[$\setminus$]p) (snd (D X)),s'>>+}}
          {\mbox{\lstinline+s [==] s'+}
            & \mbox{\lstinline+[#](fst (D X)) > 1+}
            & \mbox{\lstinline+In p (fst (D X))+}}
    \rulebreak
    \infer[\mbox{\lstinline+C_Call_Enter+}]
          {\mbox{\lstinline+<<RT_Call X ps C,s>> --[RL_Call X p,D]--> <<RT_Call X (ps[$\setminus$]p) C,s'>>+}}
          {\mbox{\lstinline+s [==] s'+}
            & \mbox{\lstinline+[#]ps > 1+}
            & \mbox{\lstinline+In p ps+}}
    \rulebreak
    \infer[\mbox{\lstinline+C_Call_Finish+}]
          {\mbox{\lstinline+<<RT_Call X ps C,s>> --[RL_Call X p,D]--> <<C,s'>>+}}
          {\mbox{\lstinline+s [==] s'+}
            & \mbox{\lstinline+[#]ps = 1+}
            & \mbox{\lstinline+In p ps+}}
  \end{gather*}
  
  \caption{Semantics of choreographies, lower layer (\lstinline+CCC_To+).}
  \label{fig:CCC_To}
\end{figure}

\begin{figure}

  \begin{gather*}
    \infer[\mbox{\lstinline+CCP_Base+}]
          {\mbox{\lstinline+(D,C,s) --[forget t]--> (D,C',s')+}}
          {\mbox{\lstinline+<<C,s>> --[t,D]--> <<C',s'>>+}}
  \end{gather*}
  
  \caption{Semantics of choreographies, middle layer (\lstinline+CCP_To+).}
  \label{fig:CCP_To}
\end{figure}

\begin{figure}

  \begin{gather*}
    \infer[\mbox{\lstinline+CCT_Base+}]
          {\mbox{\lstinline+(P,s) --[nil]-->** (P,s')+}}
          {s [==] s'}
    \qquad
    \infer[\mbox{\lstinline+CCT_Step+}]
          {\mbox{\lstinline+c1 --[t::l]-->** c3+}}
          {\mbox{\lstinline+c1 --[t]--> c2+}
            & \mbox{\lstinline+c2 --[l]-->** c3+}}
  \end{gather*}
  
  \caption{Semantics of choreographies, top layer (\lstinline+CCP_ToStar+).}
  \label{fig:CCP_ToStar}
\end{figure}

The rules defining \lstinline+CCC_To+ can be divided into three groups, which we describe in the following paragraphs.

\paragraph{Transition rules.}
Rules \lstinline+C_Com+, \lstinline+C_Sel+, \lstinline+C_Then+ and \lstinline+C_Else+ deal with execution of the first action in a choreography.

As an example, rule~\lstinline+C_Sel+ corresponds to a constructor
\begin{lstlisting}
C_Sel D p q l a C s s': s [==] s' -> CCC_To D (p --> q [l] @ a ;; C) s (RL_Sel p q l) C s'
\end{lstlisting}
Including the requirement \lstinline+s [==] s'+ instead of simply writing \lstinline+s+ in the conclusion is essential for enabling transitions between different intensional representations of the same state, which occur in practice.
In particular, confluence (discussed below) does not hold without this formulation.
The corresponding more compact rules are proved as lemmas, e.g.,
\begin{lstlisting}
Lemma C_Sel' : <<p --> q[l] @ a;; C,s>> --[RL_Sel p q l,D]--> <<C,s>>.
\end{lstlisting}
These formulations can be useful in proofs that use existential tactics to infer a previously uninstantiated target of a transition.

\paragraph{Procedure calls.}
Rules \lstinline+C_Call_Local+, \lstinline+C_Call_Start+, \lstinline+C_Call_Enter+ and \lstinline+C_Call_Finish+ allow a process to enter a procedure call, with different cases according to whether other processes have already entered the procedure and/or whether there are any other processes that still have to join it.

A procedure call is expanded when the first process joins it (rule \lstinline+C_Call_Start+).
The remaining processes and the procedure's definition are stored in a runtime term, from which we can observe transitions either by more processes entering the procedure (rule \lstinline+C_Call_Enter+) or by out-of-order execution of internal transitions of the procedure (rule \lstinline+C_Delay_Call+, discussed below).
When the last process enters the procedure, the runtime term is consumed (rule \lstinline+C_Call_Finish+).
Rule \lstinline+C_Call_Local+ addresses the edge case of a procedure that only uses one process.

\paragraph{Out-of-order execution.}
Rules \lstinline+C_Delay_Eta+, \lstinline+C_Delay_Cond+ and \lstinline+C_Delay_Call+ deal with out-of-order execution (cf.\ \cref{ex:oo-2}).
These rules require that the processes involved in the transition do not appear in the first term in the choreography; these conditions are specified by auxiliary predicates defined straightforwardly.

\begin{example}
  \label{ex:auth_chor_red}
  Consider the program \lstinline+(D,C1)+ where \lstinline+C1+ is the choreography in \Cref{ex:authentication_chor} and \lstinline+D:DefSet+ is arbitrary (there are no recursive calls in \lstinline+C1+).
  \begin{lstlisting}
(D, C1, st1) --[L_Com c ip v1]--> (D, If ip ?? (check x) Then C1t Else C1e, st2)
  \end{lstlisting}
  where \lstinline+v1 := eval_on_state Ev credentials st1 c+ is the evaluation of \lstinline+credentials+ at \lstinline+c+ in \lstinline+st1+ according to the evaluation function \lstinline+Ev+, and \lstinline+st2 := st1[[ip,x => v1]]+.
  If \lstinline+check x+ evaluates to \lstinline+true+ at \lstinline+ip+ in \lstinline+st2+, then execution continues as follows.
  \begin{lstlisting}
(D, If ip ?? (check x) Then C1t Else C1e, st2) --[L_Tau ip]--> (D, C1t, st2)
  --[L_Sel ip s left; L_Sel ip c left]-->** (D, s#token --> cdollart;; End, st2)
  --[L_Com s c v2]--> (D, End, st3)
  \end{lstlisting}
  where \lstinline+v2 := eval_on_state Ev token st2 s+ and \lstinline+st3 := st2[[c,t => v2]]+.

  If the check fails, the choreography instead continues as follows.
  \begin{lstlisting}
(D, If ip ?? (check x) Then C1t Else C1e, st2) --[L_Tau ip]--> (D, C1e, st2)
  --[L_Sel ip s right; L_Sel ip c right]-->** (D, End, st2)
  \end{lstlisting}
  In the compound transitions in the examples above, the actions in the label are executed in order.
\end{example}

\begin{example}
  Let \lstinline+Defs+ be as in \Cref{ex:file_transfer_chor} and \lstinline+C+ be the body of \lstinline+FileTransfer+.
  Consider the program \lstinline+(Defs,Call FileTransfer)+.
  The processes in the procedure \lstinline+FileTransfer+ can join it in any order as exemplified by the transitions below.

  \begin{lstlisting}
(Defs, Call FileTransfer, st) --[L_Tau c]-->
  (Defs, RT_Call FileTransfer s::nil C, st) --[L_Tau s]--> (Defs, C, st)
\end{lstlisting}

  \begin{lstlisting}
(Defs, Call FileTransfer, st) --[L_Tau s]-->
  (Defs, RT_Call FileTransfer c::nil C, st) --[L_Tau c]--> (Defs, C, st)
\end{lstlisting}
  The state \lstinline+st+ is immaterial.
\end{example}

We prove a number of useful low-level properties about transitions.
For example, we show that transitions are preserved by state equivalence.
\begin{lstlisting}
Lemma CCC_To_eq : s1 [==] s1' -> s2 [==] s2' ->
  <<C,s1>> --[tl,D]--> <<C',s2>> -> <<C,s1'>> --[tl,D]--> <<C',s2'>>.
\end{lstlisting}
This result generalises to \lstinline+CCP_To+ and \lstinline+CCP_ToStar+.
Likewise, we show that: the set of processes involved in a choreography cannot increase during execution; transitions preserve well-formedness and the set of procedure definitions; well-formed choreographies do not perform self-communications; and terminated choreographies cannot perform transitions.

\subsection{Progress, Determinism, and Confluence}
\label{sec:chor-results}

The challenging part of formalising CC is establishing the core properties of the language semantics, which are essential for more advanced results and not always proven in full detail in pen-and-paper publications.
We discuss some of the issues encountered, as these were also the driving force behind the changes relative to~\cite{CM20}.

The first key property of choreographies is that they are deadlock-free by design: any choreography that is not terminated can execute.
Since the only terminated choreography in CC is \lstinline+End+, this property also implies that any choreography either eventually reaches the terminated choreography \lstinline+End+ or runs infinitely.
\begin{lstlisting}
Theorem progress : forall P, Main P <> End -> Program_WF P ->
  forall s, exists tl c', (P,s) --[tl]--> c'.

Theorem deadlock_freedom : forall P, Program_WF P ->
  forall s ts c', (P,s) --[ts]-->** c' ->
  {Main (fst c') = End} + {exists tl c'', c' --[tl]--> c''}.
\end{lstlisting}

The second property of our semantics is that it is deterministic, in the sense that transitions can be uniquely inferred from their label or the resulting state.
These properties are essential for later results, and the need for them was the original motivation for introducing type \lstinline+RichLabel+ -- the first group of results does not hold if \lstinline+TransitionLabel+s are used in the definition of \lstinline+CCC_To+.
\begin{lstlisting}
Lemma CCC_To_deterministic : <<C,s>> --[tl1,D]--> <<C1,s1>> ->
  <<C,s>> --[tl2,D]--> <<C2,s2>> -> tl1 = tl2 -> C1 = C2 /\ s1 [==] s2.

Lemma CCC_To_deterministic_3 : <<C,s>> --[tl1,D]--> <<C',s1>> ->
  <<C,s>> --[tl2,D]--> <<C',s2>> -> tl1 = tl2.
\end{lstlisting}

The third key property is confluence, which has some relevant implications for our calculus: if a choreography has two different transition paths, then these paths either end at the same configuration, or both resulting configurations can reach the same one.
This is proved by first showing the diamond property for choreography transitions, then lifting it to one-step transitions, and finally applying induction to show it for multistep transitions.

\begin{lstlisting}
Lemma diamond_Chor :
  <<C,s>> --[tl1,D]--> <<C1,s1>> -> <<C,s>> --[tl2,D]--> <<C2,s2>> -> tl1 <> tl2 ->
  exists C' s', <<C1,s1>> --[tl2,D]--> <<C',s'>> /\ <<C2,s2>> --[tl1,D]--> <<C',s'>>.

Lemma diamond_1 : c --[ tl1 ]--> c1 -> c --[ tl2 ]--> c2 -> tl1 <> tl2 ->
  exists c', c1 --[ tl2 ]--> c' /\ c2 --[ tl1 ]--> c'.

Lemma diamond_4 : (P,s) --[ tl1 ]-->** (P1,s1) -> (P,s) --[ tl2 ]-->** (P2,s2) ->
  (exists P' tl1' tl2' s1' s2',
    (P1,s1) --[ tl1' ]-->** (P',s1') /\ (P2,s2) --[ tl2' ]-->** (P',s2') /\ s1' [==] s2').
\end{lstlisting}

As an important consequence, we get that any two executions of a choreography that end in a terminated choreography must finish in the same state.
\begin{lstlisting}
Lemma termination_unique : c --[tl1]-->** c1 -> c --[tl2]-->** c2 ->
  Main (fst c1) = End -> Main (fst c2) = End -> snd c1 [==] snd c2.
\end{lstlisting}

Using these results, we can establish Turing completeness of CC, in the sense that all of Kleene's partial recursive functions~\cite{Kleene52} can be implemented as a choreography for a suitable notion of implementation.
The structure of the proof closely follows that of~\cite{CM20}, and has been described in~\cite{CMP21a}; the interested reader is referred to those works for details.

\subsection{Discussion}
\label{sec:chor-discussion}

Formalising the proof of confluence following \cite{CM20} turned out to be a spiralling process: the pen-and-paper proof assumes some obvious properties, which were not proved; proving these required some additional lower-level lemmas; these in turn generated some even more specific lemmas; and so on.
At some point, we realised that the auxiliary lemmas accumulated already accounted for 90\% of the formalisation.
Worse, these lemmas were extremely specific and detached from the contents of \cite{CM20} -- even though we were far from done.
This led us to rethinking the design of CC, and eventually to adopting the language of~\cite{M22}.

In this section, we discuss the features of the original language that turned out to be problematic.
These regarded the handling of procedure definitions (syntax) and the treatment of procedure calls and out-of-order execution (semantics).

\paragraph{Syntax.}
Procedures were initially defined by including a term \lstinline+def X=CX in C+ in the grammar defining choreographies.
While this removed the need for a separate notion of program, it introduced several dimensions of complexity.
Even the notion of terminated choreography was nontrivial, since \lstinline+End+ could occur arbitrarily deep inside some of these terms.
This made it hard to ensure that the Coq definition was an adequate representation of the informal notion in \cite{CM20}, affecting all results regarding termination, progress, and deadlock-freedom.
With the current syntax, terminated programs are exactly those whose for which \lstinline+Main P=End+.

Additionally, the name \lstinline+X+ in \lstinline+def X=CX in C+ acts as a binder, which added all the usual problems of working with binders -- in particular, having to deal with capture-avoiding substitutions and $\alpha$-renaming.
In the current language, procedure names are statically determined and fixed, so there is no need to rename them ever, and they can be treated as constants.
This constructor also allowed for unintuitive choreographies, e.g., \lstinline+def X=CX in C+ where the choreography \lstinline+CX+ itself contains additional procedure definitions.

Pairing procedure definitions with choreographies in programs yields a cleaner theory, and the overhead of an additional layer is a very small price to pay for the simplicity gained.
This approach had been proposed earlier~\cite{CM17}, and the two formulations are argued to be equally expressive in~\cite{CLM17}.

\paragraph{Semantics.}
Instead of a labelled transition system, the semantics of~\cite{CM20} was a reduction semantics that used a structural precongruence relation to model out-of-order execution and to unfold procedure definitions.

To understand this issue, consider again \cref{ex:oo-2}, which shows a choreography that has two possible initial transitions.
In a framework with reductions and structural precongruence, the out-of-order transition is obtained by first rewriting the choreography as $\com{\pid o_2}{\pid s_2}; \com{\pid o_1}{\pid s_1}$ and then applying rule \rname{Com}.
The set of legal rewritings is formally defined by the structural precongruence relation $\preceq$, and there is a rule in the semantics that closes the transition relation under it.

In any proofs about the semantics, an approach using structural precongruence needs to take into account all the possible ways into which choreographies may be rewritten in a reduction.
Concretely, in the proof of confluence, where there are two reductions, there are four possible places where choreographies are rewritten; given the high number of rules defining structural precongruence, this led to an explosion of the number of cases.
Furthermore, induction hypotheses typically were not strong enough, requiring us to resort to complicated auxiliary notions such as explicitly measuring the size of the derivation of transitions, and proving that rewritings could be normalised.
This process led to a seemingly ever-growing number of auxiliary lemmas that needed to be proved, with no counterpart in the original reference~\cite{CM20}, and after several months of work with little progress it became evident that the problem lay in the formalism.

\paragraph{Summary.}
The current proof of confluence takes about 300 lines of Coq code, including a total of 11 lemmas.
This is in stark contrast with the previous attempt, which while still unfinished already included over 30 lemmas with extremely long proofs.

With the current definitions, the theory of CC is formalised in two files.
The first file, which defines the preliminaries, contains 24 definitions, 60 lemmas and around 740 lines of code.
The second file, defining the syntax and semantics of CC and proving properties about it (including all the ones described herein), contains 32 definitions, 126 lemmas, 2 theorems and around 2300 lines of code.

\section{The process language}
\label{sec:sp}

The second part of our formalisation concerns the process calculus that we use for implementing CC: Stateful Processes (SP).
We follow the pen-and-paper design presented in \cite{M22}.
SP is used to define networks of processes running in parallel, each with its own behaviour, that can interact by direct messaging.

\subsection{Syntax}
\label{sec:sp-syntax}

The syntax of SP is structured in three layers: behaviours, which express the local actions performed by individual processes; networks, which combine processes in a system where they can interact; and programs, which pair a network with a set of procedure definitions (which all processes can call).
As with CC, we assume an underlying signature.

The constructors for behaviours correspond to those for choreographies, but interactions are now split between the two different roles involved (sender and receiver).
The type \coc{Behaviour} is defined inductively from the grammar below.

\begin{lstlisting}
 B $\Coloneqq$ End $\mid$ p!e @! a; B $\mid$ p?x @? a; B $\mid$ p(+)l @+ a; B $\mid$ p & mB1 // mB2
      $\mid$ If e Then B1 Else B2 $\mid$ Call X
mB $\Coloneqq$ None $\mid$ Some (a,B)
\end{lstlisting}
Conditionals, procedure calls, and the terminated behaviour are standard and similar to the corresponding constructs in CC.

A term \coc{p!e @! a} represents a send action towards \coc{p}, where \coc{e} is the expression used to compute the value to be sent and \coc{a} is an annotation.
Dually, a term \coc{p?x @? a} represents a receive action where a value received from \coc{p} is stored in the local variable \coc{x} (\coc{a} is, again, an annotation).

A selection action \lstinline!p(+)l @+ a; B! is similar to a send action (label \coc{l} is sent to \coc{p}).
The dual action needs to offer a behaviour for \coc{l}, but may also accept other labels.
In pen-and-paper presentations, these \emph{branching terms} are typically defined as partial functions from labels to behaviours.

Formalising this informal description is challenging.
A natural choice would be to include a constructor \lstinline+Branching : Pid -> (Label -> option Behaviour) -> Behaviour+.
However, this is problematic for defining EPP, which relies on a recursively defined function on pairs of behaviours called \emph{merging} (cf.\ \cref{sec:merge}).
Defining this function directly in Coq is unwieldy because of the complexity of writing the appropriate term of type \coc{Label -> option Behaviour} given the corresponding subterms from the arguments.

Using partial functions also seems like an overkill, considering that there are only two possible labels.
Instead, we include a constructor
\begin{lstlisting}
Branching : Pid -> option (Ann*Behaviour) -> option (Ann*Behaviour) -> Behaviour
\end{lstlisting}
that registers explicitly the behaviours offered for each of the two possible labels, in order.
This design choice avoids the aforementioned issues, at the cost of making our development harder to generalise to larger sets of labels in the future.

Because of the option types in \lstinline+Branching+, the induction principles generated automatically for \lstinline+Behaviour+ are not strong enough (they do not include induction hypotheses over the \lstinline+Behaviour+s appearing within branching terms).
To overcome this, we define an auxiliary function \lstinline+depth:Behaviour -> nat+ measuring the depth of the AST corresponding to a \lstinline+Behaviour+, use it to prove the expected general induction principle, and define a tactic \lstinline+BInduction B+ that applies it.

\paragraph{Networks.}
Networks are simply (total) functions from processes to behaviours.
\begin{lstlisting}
Definition Network := Pid -> Behaviour.
\end{lstlisting}
We define extensional equality of networks \lstinline+N (==) N'+ in the expected way and show that it is an equivalence relation.
We support the common notation for writing networks by including a function for constructing singleton networks \lstinline+p[B]+, a parallel composition operator \lstinline+N | N'+, and a removal operator \lstinline+N \ p+ (recall the description in \cref{sec:background}).

For simplicity, we do not require disjoint support in parallel composition: if both networks define a nonterminated behaviour for \lstinline+p+, the result of \lstinline+(N | N') p+ and \lstinline+(N' | N) p+ is different.
Although this may seem odd, it has the advantage of making parallel composition total.
We show that parallel composition is commutative under the assumption that the two composed networks have disjoint supports.
\begin{lstlisting}
Lemma Behaviour_eq_End_dec : forall (b:Behaviour), {b=End} + {b<>End}.

(* N | N' stands for (Par N N') *)
Definition Par (N N':Network) :=
  fun p => if (Behaviour_eq_End_dec (N p)) then N' p else N p.

Definition Network_disjoint (N N':Network) := forall p, N p = End \/ N' p = End.

Lemma Par_comm : Network_disjoint N N' -> (N | N') (==) (N' | N).
\end{lstlisting}

Our library includes a number of results to reason about the network operations, including very specific lemmas dealing with networks that appear in the rules defining the semantics of SP, e.g., that updating the behaviours of two distinct processes yields the same result independent of the order of the updates.

\paragraph{Programs and well-formedness.}
As before, a program is a pair consisting of a set of procedure definitions and a network.
\begin{lstlisting}
Definition DefSetB := RecVar -> Behaviour.
Definition Program := DefSetB * Network.
\end{lstlisting}

Well-formedness is significantly simpler than for choreographies.
If \lstinline+B:Behaviour+, then \lstinline+B+ is well-formed, \lstinline+Behaviour_WF B+, as long as no process in \lstinline+B+ attempts to communicate with itself.
\lstinline+N:Network+ is well-formed, \lstinline+Network_WF N+, if all processes are mapped to well-formed behaviours.
This is not decidable in general, but it is under the assumption that all processes outside a given set \lstinline+ps+ are mapped to \lstinline+End+ -- an assumption that holds for all networks that can be written explicitly using parallel composition of singleton networks.

Well-formedness of programs does not make sense: well-formedness of a behaviour depends on who is executing it, but a procedure definition has no information about which processes will call it.

\begin{example}\label{ex:authentication_net}
Consider the network \lstinline+N = c[Bc] | s[Bs] | ip[Bip]+, where:
\begin{lstlisting}
Bc := ip!credentials; ip & Some (s?t; End) // Some End
Bs := ip & Some (c!token; End) // Some End

Bip := c?x; Bip'
Bip' := If (check x) Then (s(+)left; c(+)left; End) Else (s(+)right; c(+)right; End)
\end{lstlisting}
This network implements the choreography in Example~\ref{ex:authentication_chor}.
\end{example}

\subsection{Semantics}
\label{sec:sp-semantics}

The semantics of SP is again defined by a labelled transition system.
Transitions for communications match dual actions in two processes, while conditionals and procedure calls simply run locally.
There are again three layers of definitions, which are shown in \Cref{fig:SP_To,fig:SPP_To,fig:SPP_ToStar}, and two types of transition labels (as in CC).
Transitions support suggestive notations: \lstinline"<<N,s>> --[tl,D]--> <<N',s'>>" for \lstinline+(SP_To D N s tl N' s')+, \lstinline"C --[l]--> C'" for \lstinline+(SPP_To C l C')+, and \lstinline"C --[ls]-->** C'" for \lstinline+(SPP_ToStar C ls C')+.

\begin{figure}
  \begin{gather*}
    \infer[\mbox{\lstinline+S_Com+}]
          {\mbox{\lstinline+<<N,s>> --[RL_Com p v q x,D]--> <<N',s'>>+}}
          {\begin{array}c
              \mbox{\lstinline+N p = (q ! e @!a ; B)+} \\
              \mbox{\lstinline+N q = (p ? x @? a'; B')+}
            \end{array}
            &
            \begin{array}c
              \mbox{\lstinline+v := eval_on_state Ev e s p+} \\
              \mbox{\lstinline+N' (==) (N \ p \ q | p[B] | q[B'])+}
            \end{array}
            & \mbox{\lstinline+s' [==] (s[[q,x => v]])+}}
    \rulebreak
    \infer[\mbox{\lstinline+S_LSel+}]
          {\mbox{\lstinline+<<N,s>> --[RL_Sel p q left,D]--> <<N',s'>>+}}
          {\begin{array}c
              \mbox{\lstinline"N p = (q (+) left @+ a ; B)"} \\
              \mbox{\lstinline+N q = (p & Some (a',Bl) // Br)+}
            \end{array}
            & \mbox{\lstinline+N' (==) (N \ p \ q | p[B] | q[Bl])+}
            & \mbox{\lstinline+s [==] s'+}}
    \rulebreak
    \infer[\mbox{\lstinline+S_RSel+}]
          {\mbox{\lstinline+<<N,s>> --[RL_Sel p q right,D]--> <<N',s'>>+}}
          {\begin{array}c
              \mbox{\lstinline"N p = (q (+) right @+ a ; B)"} \\
              \mbox{\lstinline+N q = (p & Bl // Some (a',Br))+}
            \end{array}
            & \mbox{\lstinline+N' (==) (N \ p \ q | p[B] | q[Br])+}
            & \mbox{\lstinline+s [==] s'+}}
    \rulebreak
    \infer[\mbox{\lstinline+S_Then+}]
          {\mbox{\lstinline+<<N,s>> --[RL_Cond p,D]--> <<N',s'>>+}}
          {\begin{array}c
              \mbox{\lstinline+N p = (If b Then B1 Else B2)+} \\
              \mbox{\lstinline+eval_on_state BEv b s p = true+}
            \end{array}
            & \mbox{\lstinline+N' (==) (N \ p | p[B1])+}
            & \mbox{\lstinline+s [==] s'+}}
    \rulebreak
    \infer[\mbox{\lstinline+S_Else+}]
          {\mbox{\lstinline+<<N,s>> --[RL_Cond p,D]--> <<N',s'>>+}}
          {\begin{array}c
              \mbox{\lstinline+N p = (If b Then B1 Else B2)+} \\
              \mbox{\lstinline+eval_on_state BEv b s p = false+}
            \end{array}
            & \mbox{\lstinline+N' (==) (N \ p | p[B2])+}
            & \mbox{\lstinline+s [==] s'+}}
    \rulebreak
    \infer[\mbox{\lstinline+S_Call+}]
          {\mbox{\lstinline+<<N,s>> --[RL_Call X p,D]--> <<N',s'>>+}}
          {\mbox{\lstinline+N p = Call X+}
            & \mbox{\lstinline+N' (==) (N \ p | p[D X])+}
            & \mbox{\lstinline+s [==] s'+}}
  \end{gather*}
  \caption{Semantics of networks, bottom layer (\lstinline+SP_To+).}
  \label{fig:SP_To}
\end{figure}

\begin{figure}
  \begin{gather*}
    \infer[\mbox{\lstinline+SPP_Base+}]
          {\mbox{\lstinline+(D,N,s) --[forget t]--> (D,N',s')+}}
          {\mbox{\lstinline+<<N,s>> --[t,D]--> <<N',s'>>+}}
  \end{gather*}
  \caption{Semantics of networks, middle layer (\lstinline+SPP_To+).}
  \label{fig:SPP_To}
\end{figure}

\begin{figure}
  \begin{gather*}
    \infer[\mbox{\lstinline+SPT_Base+}]
          {\mbox{\lstinline+(P,s) --[nil]-->** (P,s')+}}
          {\mbox{\lstinline+s [==] s'+}}
    \qquad
    \infer[\mbox{\lstinline+SPT_Step+}]
          {\mbox{\lstinline+c1 --[t::l]-->** c3+}}
          {\mbox{\lstinline+c1 --[t]--> c2+}
            & \mbox{\lstinline+c2 --[l]-->** c3+}}
  \end{gather*}
  \caption{Semantics of networks, top layer (\lstinline+SPP_ToStar+).}
  \label{fig:SPP_ToStar}
\end{figure}

These definitions warrant similar observations as those for the semantics of CC.
Transitions include premises on network equality and state equality, rather than requiring specific values.
We include some lemmas stating the more restricted rules, both as a sanity check and because they can be useful to instantiate variables created by the use of existential tactics in proofs.
\begin{lstlisting}
Lemma S_LSel' : N p = (q (+) left @+ a ; B) -> N q = (p & Some (a',Bl) // Br) ->
  <<N,s>> --[RL_Sel p q left,D]--> <<N \ p \ q | p[B] | q[Bl],s>>.
\end{lstlisting}

There are two rules for reducing selections, one for each label.
This is a deviation for standard practice (where there is a single rule and a premise matching the label in both behaviours) stemming from our design choice of avoiding functions in branching terms.
Having an extra rule generates additional cases in induction proofs, but this formulation effectively simplifies the formalisation by eliminating one layer of inversion.

\begin{example}
  We illustrate the possible transitions of the network from Example~\ref{ex:authentication_net}.
  We abbreviate the behaviours of processes that do not change in a reduction to \lstinline+...+ to make it clearer what parts of the network are changed.
  Furthermore, we omit trailing \lstinline+End+s in \lstinline+Behaviour+s.

  The network starts by performing the transition
\begin{lstlisting}
(D, c[Bc] | s[Bs] | ip[Bip], st1) --[L_Com c ip v1]-->
  (D, c[ip & Some (s?t) // Some End] | s[...] | ip[Bip'], st2)
\end{lstlisting}
where \lstinline+v1+ and \lstinline+st2+ are as in Example~\ref{ex:auth_chor_red}.

If \lstinline+eval_on_state (check x) st2 ip=true+, execution continues as
\begin{lstlisting}
(D, c[ip & Some (s?t) // Some End] | s[Bs] | ip[Bip'], st2)
  --[L_Tau ip]-->             (D, c[...] | s[...] | ip[s(+)left;c(+)left], st2)
  --[L_Sel ip s left]--> (D, c[...] | s[c!token] | ip[c(+)left], st2)
  --[L_Sel ip c left]--> (D, c[s?t] | s[...] | ip[End], st2)
  --[L_Com s c v2]-->         (D, c[End] | s[End] | ip[End], st3)
\end{lstlisting}
where \lstinline+v2+ and \lstinline+st3+ are again as in Example~\ref{ex:auth_chor_red}.
Otherwise, it continues as follows.
\begin{lstlisting}
(D; c[ip & Some (s?t) // Some End] | s[Bs] | ip[Bip'], st2) 
  --[L_Tau ip]-->              (D; c[...] | s[...] | ip[s(+)right;c(+)right], st2)
  --[L_Sel ip s right]--> (D; c[...] | s[End] | ip[c(+)right], st2)
  --[L_Sel ip c right]--> (D; c[End] | s[End] | ip[End], st2)
\end{lstlisting}
The labels in these reductions are exactly as in Example~\ref{ex:auth_chor_red}.
\end{example}

\subsection{Determinism and confluence}
\label{sec:sp-results}
As for CC, we prove a number of useful results about the semantics of SP.
These can be roughly divided in two groups: results showing that reductions are stable under the extensional equalities on the different types involved, and properties on the actual transitions.
While the results in the first category are not surprising, they are useful and show that the definitions make sense.
\begin{lstlisting}
Lemma SP_To_eq : s1 [==] s1' -> s2 [==] s2' ->
  <<N,s1>> --[tl,D]--> <<N',s2>> -> <<N,s1'>> --[tl,D]--> <<N',s2'>>.

Lemma SP_To_Network_eq : N1 (==) N2 ->
  <<N1,s>> --[tl,D]--> <<N',s'>> -> <<N2,s>> --[tl,D]--> <<N',s'>>.

Lemma SP_To_Defs_wd : (forall X, D X = D' X) ->
  <<N,s>> --[tl,D]--> <<N',s'>> -> <<N,s>> --[tl,D']--> <<N',s'>>.
\end{lstlisting}

While determinism and confluence are similar to the corresponding results to CC, they are not as interesting: for networks generated by EPP (which are the ones we are interested in), these results would follow by the same properties for choreographies.

The formalisation of SP consists of 25 definitions, 81 lemmas, 11 simple tactics, and approximately 1960 lines of Coq code.

\section{Endpoint projection}
\label{sec:epp}

As with the simple language from \Cref{sec:background}, the intuition for generating process implementations is that each choreographic action should be projected to the corresponding process action.
The prototypical example is the value communication \lstinline+p#e --> qdollarx @a+, which should be projected to a send action \lstinline+q!e @!a+ for \lstinline+p+, to a receive action \lstinline+p?x @?a+ for \lstinline+q+, and skipped for any other processes.

In the presence of conditionals, this intuition is not enough.
Projecting a conditional \lstinline+If p.b Then Ct Else Ce+ for any process other than \lstinline+p+, say \lstinline+q+, is nontrivial, because \lstinline+q+ has no way of knowing which branch should be executed.
Therefore \lstinline+q+'s behaviour must combine the projections obtained for \lstinline+Ct+ and \lstinline+Ce+.

This problem is commonly known as \emph{knowledge of choice}, and one of the solutions relies on the usage of label selections \cite{CDP11,CHY12}.
If \lstinline+q+'s behaviour should depend on the result of \lstinline+p+'s local evaluation, then the result of this evaluation should be communicated to \lstinline+q+ by means of a label selection.
The two possible behaviours can then be combined in a branching term offering two different options.

\subsection{Merge}
\label{sec:merge}

A standard way of combining behaviours to solve the problem above is the \emph{merge} operator~\cite{CHY12}: a partial binary operator that returns a behaviour combining all possible executions of its arguments (if possible).
In SP, two behaviours can be merged only if they are built from the same constructor with matching parameters.
So if \lstinline+B1+ can be merged with \lstinline+B2+ to yield \lstinline+B+, we can also merge \lstinline+p?x @? a; B1+ with \lstinline+p?x @? a; B2+ to obtain \lstinline+p?x @? a; B+, but \lstinline+p?x @? a; B1+ can never be merged with \lstinline+q?x @? a; B2+ for \lstinline+p<>q+ (different arguments) or with \lstinline+q!e @! a; B2+ (different constructor).

The only exception is branching terms, where merge can combine offers on different labels.
For example, merging \lstinline+p & Some (a,B) // None+ with \lstinline+p & None // Some (a',B')+ yields \lstinline+p & Some (a,B) // Some (a',B')+.
In this way, the prototypical choreographic conditional \lstinline+If p??b Then (p-->q[left];; q.e --> p;; End) Else (p-->q[right];; End)+ can be projected for \lstinline+q+ as \lstinline+p & Some (p!e; End) // Some End+.

The partiality of merge again poses a formalisation problem.
Our original approach~\cite{CMP21b} defined an auxiliary type \lstinline+XBehaviour+ that extends the syntax of behaviours with a constructor \lstinline+XUndefined : XBehaviour+.
In this work, instead, we define a ternary relation \lstinline+merge : Behaviour -> Behaviour -> Behaviour -> Prop+.\footnote{Technically, because \lstinline+merge+ is defined on a separate file, all types defined in the formalisation of SP need the signature as a parameter. Since this signature is fixed, we omit it everywhere.}
While this design requires two additional lemmas stating that this relation is functional and computable, it significantly simplified this part of the formalisation (both in size and complexity of the proofs).
As an example, \cite{CMP21b} reported a number of inversion results, e.g., if merging two behaviours yields a behaviour starting with a send action, then both arguments start with that same action.
All these results can now be obtained directly by applying inversion on the relevant hypotheses.

The full definition of \lstinline+merge+ includes 22 clauses.
\Cref{fig:merge} lists all representative cases; the missing clauses deal with the remaining combinations of \lstinline+None+ / \lstinline+Some+ subterms in branching terms (see \Cref{sec:epp-discussion} for a discussion on the exponential dependency of the number of clauses on the number of labels).
We also define the suggestive notation \lstinline+B1 [V] B2 == B+ for \lstinline+merge B1 B2 B+, which reminds us that \lstinline+merge+ is a partial function.
\begin{figure}
  \begin{gather*}
    \infer[\mbox{\lstinline+merge_End+}]
          {\mbox{\lstinline+End [V] End == End+}}
          {}
    \quad
    \infer[\mbox{\lstinline+merge_Send+}]
          {\mbox{\lstinline+p!e @!a; B1 [V] p!e @!a; B2 == p!e @!a; B+}}
          {\mbox{\lstinline+B1 [V] B2 == B+}}
    \rulebreak
    \infer[\mbox{\lstinline+merge_Recv+}]
          {\mbox{\lstinline+p?x @?a; B1 [V] p?x @?a; B2 == p?x @?a; B+}}
          {\mbox{\lstinline+B1 [V] B2 == B+}}
    \rulebreak
    \infer[\mbox{\lstinline+merge_Sel+}]
          {\mbox{\lstinline"p(+)l @+a; B1 [V] p(+)l @+a; B2 == p(+)l @+a; B"}}
          {\mbox{\lstinline+B1 [V] B2 == B+}}
    \rulebreak
    \infer[\mbox{\lstinline+merge_Branching_NNNN+}]
          {\mbox{\lstinline+p & None // None [V] p & None // None == p & None // None+}}
          {}
    \rulebreak
    \infer[\mbox{\lstinline+merge_Branching_SNNN+}]
          {\begin{multlined}
            \coc{p & Some (aL,bL) // None [V] p & None // None}
            \\
            \coc{== p & Some (aL,bL) // None}
          \end{multlined}
          }
          {}
    \rulebreak
    \infer[\mbox{\lstinline+merge_Branching_SNSN+}]
          {
            \begin{multlined}
              \coc{p & Some (aL,bL1) // None [V] p & Some(aL,bL2) // None}
              \\
              \coc{== p & Some (aL,bL) // None}
            \end{multlined}
          }
          {\mbox{\lstinline+bL1 [V] bL2 == bL+}}
    \rulebreak
    \infer[\mbox{\lstinline+merge_Branching_SSSN+}]
          {\begin{multlined}
            \coc{p & Some (aL,bL1) // Some (aR,bR) [V] p & Some(aL,bL2) // None} \\
            \coc{== p & Some (aL,bL) // Some (aR,bR)}
            \end{multlined}
          }
          {\mbox{\lstinline+bL1 [V] bL2 == bL+}}
    \rulebreak
    \infer[\mbox{\lstinline+merge_Branching_SSSS+}]
          {\begin{multlined}
            \coc{p & Some (aL,bL1) // Some (aR,bR1)}
            \\
            \coc{[V] p & Some(aL,bL2) // Some(aR,bR2)}
            \\
            \coc{== p & Some (aL,bL) // Some (aR,bR)}
            \end{multlined}
          }
          {\mbox{\lstinline+bL1 [V] bL2 == bL+}
            & \mbox{\lstinline+bR1 [V] bR2 == bR+}}
    \rulebreak
    \infer[\mbox{\lstinline+merge_Cond+}]
          {\begin{multlined}
            \coc{If p??e Then Bt1 Else Bt2 [V] If p??e Then Be1 Else Be2}
            \\
            \coc{== If p??e Then Bt Else Be}
            \end{multlined}
          }
          {\mbox{\lstinline+Bt1 [V] Bt2 == Bt+}
            & \mbox{\lstinline+Be1 [V] Be2 == Be+}}
    \rulebreak
    \infer[\mbox{\lstinline+merge_Call+}]
          {\mbox{\lstinline+Call X [V] Call X == Call X+}}
          {}
  \end{gather*}
  \caption{Definition of the merge relation.}
  \label{fig:merge}
\end{figure}

We show that merge is functional, decidable, and preserves well-formedness.
\begin{lstlisting}
Lemma merge_unique : B1 [V] B2 == B -> B1 [V] B2 == B' -> B = B'.
Lemma merge_dec : { B | B1 [V] B2 == B } + { ~exists B, B1 [V] B2 == B }.
Lemma merge_WF : B1 [V] B2 == B ->
  Behaviour_WF _ p B1 -> Behaviour_WF _ p B2 -> Behaviour_WF _ p B.
\end{lstlisting}
Decidability is formulated using the stronger existential quantifier so that we can also obtain the existential witness to use in further definitions.

\subsection{Branching order}
\label{sec:pruning}

In the literature, the arguments of merge and its result, when defined, are in a relation known as the \emph{branching order} \cite{CHY12,M22}.
This is formalised as yet another inductive type, defined by the rules in \Cref{fig:more-branches}.\footnote{In the formalisation, these definitions precede that of \lstinline+merge+. This was chosen because the branching relation is more primitive than the notion of merging. For presentation purposes, though, it makes more sense to invert this order.}
We call the relation \lstinline+more_branches+, for which we define the infix notation \lstinline+[>>]+.

\begin{figure}
  \begin{gather*}
    \infer[\mbox{\lstinline+MB_End+}]
          {\mbox{\lstinline+End [>>] End+}}
          {}
    \qquad
    \infer[\mbox{\lstinline+MB_Send+}]
          {\mbox{\lstinline+p!e @! a; B [>>] p!e @! a; B'+}}
          {\mbox{\lstinline+B [>>] B'+}}
    \rulebreak
    \infer[\mbox{\lstinline+MB_Recv+}]
          {\mbox{\lstinline+p?x @? a; B [>>] p?x @? a; B'+}}
          {\mbox{\lstinline+B [>>] B'+}}
    \qquad
    \infer[\mbox{\lstinline+MB_Sel+}]
          {\mbox{\lstinline"p(+)l @+ a; B [>>] p(+)l @+ a; B'"}}
          {\mbox{\lstinline+B [>>] B'+}}
    \rulebreak
    \infer[\mbox{\lstinline+MB_Branching_NN+}]
          {\mbox{\lstinline+p & mBl // mBr [>>] p & None // None+}}
          {}
    \rulebreak
    \infer[\mbox{\lstinline+MB_Branching_NS+}]
          {\mbox{\lstinline+p & mBl // Some (a,Br) [>>] p & None // Some (a,Br')+}}
          {\mbox{\lstinline+Br [>>] Br'+}}
    \rulebreak
    \infer[\mbox{\lstinline+MB_Branching_SN+}]
          {\mbox{\lstinline+p & Some (a,Bl) // mbR [>>] p & Some (a,Bl') // None+}}
          {\mbox{\lstinline+Bl [>>] Bl'+}}
    \rulebreak
    \infer[\mbox{\lstinline+MB_Branching_SS+}]
          {\mbox{\lstinline+p & Some (a,Bl) // Some (a',Br) [>>] p & Some (a,Bl') // Some (a',Br')+}}
          {\mbox{\lstinline+Bl [>>] Bl'+}
            & \mbox{\lstinline+Br [>>] Br'+}}
    \rulebreak
    \infer[\mbox{\lstinline+MB_Cond+}]
          {\mbox{\lstinline+If p Then B1 Else B2 [>>] If p Then B1' Else B2'+}}
          {\mbox{\lstinline+B1 [>>] B1'+}
            & \mbox{\lstinline+B2 [>>] B2'+}}
    \qquad
    \infer[\mbox{\lstinline+MB_Call+}]
          {\mbox{\lstinline+Call X [>>] Call X+}}
          {}
  \end{gather*}
  \caption{Definition of the branching order.}
  \label{fig:more-branches}
\end{figure}

The branching order is reflexive, transitive and antisymmetric.
It is pointwise extended to networks by defining \lstinline+ N (>>) N' := forall p, N p [>>] N' p+ where \lstinline+(>>)+ is infix notation for \lstinline+more_branches_N : Network -> Network -> Prop+.
This relation is again reflexive, transitive and antisymmetric (with respect to extensional equality).

More interestingly, adding branches to some behaviours in a network does not eliminate any transitions that the network can do.
\begin{lstlisting}
Lemma SP_To_MBN : <<N1,s>> --[tl,D]--> <<N2,s'>> -> N1' (>>) N1 ->
  (forall X, D X = D' X) -> exists N2', <<N1',s>> --[tl,D']--> <<N2',s'>> /\ N2' (>>) N2.
\end{lstlisting}
(The quantification on \lstinline+D'+ makes this lemma easier to apply.)

We can now justify the notation for \lstinline+merge+: it is the partial join  for the branching order, in the sense that if two behaviours have an upper bound, then they are mergeable and their merging is their least upper bound.
\begin{lstlisting}
Lemma MB_merge : B1 [>>] B2 <-> B1 [V] B2 == B1.

Lemma merge_is_upper_bound : B1 [V] B2 == B -> B [>>] B1.

Lemma MB_has_lub : B [>>] B1 -> B [>>] B2 -> exists B', B1 [V] B2 == B' /\ B [>>] B'.

Lemma merge_is_lub : B [>>] B1 -> B [>>] B2 -> forall B', B1 [V] B2 == B' -> B [>>] B'.
\end{lstlisting}

Another key result is that the branching order is stable under merging:
\begin{lstlisting}
Lemma MB_yields_merge : B1 [>>] B1' -> B2 [>>] B2' -> B1 [V] B2 == B ->
  exists B', B1' [V] B2' == B' /\ B [>>] B'.
\end{lstlisting}
As we will see, this result is essential for the cases of the EPP theorem dealing with conditionals.

Lastly, we prove the algebraic properties of \lstinline+merge+ -- idempotency, commutativity, and associativity -- by exploiting the relationship between \lstinline+merge+ and the branching order.

\subsection{Projection}
\label{sec:epp-projection}

We can now define the projection of a choreography for an individual process.
Since this definition relies on \lstinline+merge+ for the case of the conditionals, it is also a partial function.
We define it inductively as a relation
\begin{lstlisting}
bproj : DefSet -> Choreography Sig -> Pid -> Behaviour Sig' -> Prop
\end{lstlisting}
and abbreviate \lstinline+bproj D C p B+ to \lstinline+[[D,C | p]] == B+.

The type of \lstinline+bproj+ also reveals a new feature of projection when compared to the simple language from \Cref{sec:background}: the signature for the target instance of SP is different than that of the source instance of CC.
The reason for this lies in the presence of procedure definitions: each procedure yields several projected procedures, one for each process in the choreography.\footnote{More precisely, one nontrivial procedure for each process actually involved in it -- the remaining ones are all \lstinline+End+.}
The type of procedure names in the target of projection is thus \lstinline+RecVar*Pid+; this can be seen in the rule for projecting procedure calls, which is included with the remaining rules in \Cref{fig:bproj}.

\begin{figure}
  \begin{gather*}
    \infer[\mbox{\lstinline+bproj_End+}]
          {\mbox{\lstinline+[[D,End | p]] == End+}}
          {}
    \rulebreak
    \infer[\mbox{\lstinline+bproj_Send+}]
          {\mbox{\lstinline+[[D,p#e --> qdollarx @a ;; C | p]] == q!e @!a; B+}}
          {\mbox{\lstinline+[[D,C | p]] == B+}}
    \rulebreak
    \infer[\mbox{\lstinline+bproj_Recv+}]
          {\mbox{\lstinline+[[D,p#e --> qdollarx @a ;; C | q]] == p?x @?a; B+}}
          {\mbox{\lstinline+p<>q+}
            & \mbox{\lstinline+[[D,C | q]] == B+}}
    \rulebreak
    \infer[\mbox{\lstinline+bproj_Com+}]
          {\mbox{\lstinline+[[D,p#e --> qdollarx @a ;; C | r]] == B+}}
          {\mbox{\lstinline+p<>r+}
            & \mbox{\lstinline+q<>r+}
            & \mbox{\lstinline+[[D,C | r]] == B+}}
    \rulebreak
    \infer[\mbox{\lstinline+bproj_Pick+}]
          {\mbox{\lstinline"[[D,p --> q[l] @a ;; C | p]] == q(+)l @+a; B"}}
          {\mbox{\lstinline+[[D,C | p]] == B+}}
    \rulebreak
    \infer[\mbox{\lstinline+bproj_Left+}]
          {\mbox{\lstinline+[[D,p --> q[left] @a ;; C | q]] == p & Some (a,B) // None+}}
          {\mbox{\lstinline+p<>q+}
            & \mbox{\lstinline+[[D,C | q]] == B+}}
    \rulebreak
    \infer[\mbox{\lstinline+bproj_Right+}]
          {\mbox{\lstinline+[[D,p --> q[right] @a ;; C | q]] == p & None // Some (a,B)+}}
          {\mbox{\lstinline+p<>q+}
            & \mbox{\lstinline+[[D,C | q]] == B+}}
    \rulebreak
    \infer[\mbox{\lstinline+bproj_Sel+}]
          {\mbox{\lstinline+[[D,p --> q[l] @a ;; C | r]] == B+}}
          {\mbox{\lstinline+p<>r+}
            & \mbox{\lstinline+q<>r+}
            & \mbox{\lstinline+[[D,C | r]] == B+}}
    \rulebreak
    \infer[\mbox{\lstinline+bproj_Cond+}]
          {\mbox{\lstinline+[[D,If p??b Then C1 Else C2 | p]] == If b Then B1 Else B2+}}
          {\mbox{\lstinline+[[D,C1 | p]] == B1+}
            & \mbox{\lstinline+[[D,C2 | p]] == B2+}}
    \rulebreak
    \infer[\mbox{\lstinline+bproj_Cond'+}]
          {\mbox{\lstinline+[[D,If p??b Then C1 Else C2 | p]] == B+}}
          {\mbox{\lstinline+p<>r+}
            & \mbox{\lstinline+[[D,C1 | p]] == B1+}
            & \mbox{\lstinline+[[D,C2 | p]] == B2+}
            & \mbox{\lstinline+B1 [V] B2 == B+}}
    \rulebreak
    \infer[\mbox{\lstinline+bproj_Call_in+}]
          {\mbox{\lstinline+[[D,Call X | p]] == Call (X,p)+}}
          {\mbox{\lstinline+In p (fst (D X))+}}
    \rulebreak
    \infer[\mbox{\lstinline+bproj_Call_out+}]
          {\mbox{\lstinline+[[D,Call X | p]] == End+}}
          {\coc{$\sim$In p (fst (D X))}}
    \rulebreak
    \infer[\mbox{\lstinline+bproj_RT_Call_in+}]
          {\mbox{\lstinline+[[D,RT_Call X ps C | p]] == Call (X,p)+}}
          {\mbox{\lstinline+In p ps+}}
    \rulebreak
    \infer[\mbox{\lstinline+bproj_Call_out+}]
          {\mbox{\lstinline+[[D,RT_Call X ps C | p]] == B+}}
          {\mbox{\lstinline+$\sim$In p ps+}
            & \mbox{\lstinline+[[D,C | p]] == B+}}
  \end{gather*}
  \caption{Rules for projecting a choreography for a given target process. The notations are the ones printed by Coq, but they are not parsable due to the the different signatures.}
  \label{fig:bproj}
\end{figure}

We show that \lstinline+bproj+ is functional and decidable, and that it returns well-formed behaviours for choreographies without self-communications.

From \coc{bproj} we obtain several notions of projectability: relative to a process or a set of processes, and projectability of \lstinline+D:DefSet+ -- which requires each procedure to be projectable relative to its set of used processes.
\begin{lstlisting}
Definition projectable_B D C p := exists B, [[D,C | p]] == B.
Definition projectable_C D C ps := Forall (fun p => projectable_B D C p) ps.
Definition projectable_D D := forall X, projectable_C D (snd (D X)) (fst (D X)).
\end{lstlisting}

A program is projectable if the main choreography is projectable for all its processes and the set of procedure definitions is projectable.

\begin{lstlisting}
Definition projectable_P P :=
  projectable_C (Procedures P) (Main P) (CCP_pn P) /\
  projectable_D (Procedures P).
\end{lstlisting}

Finally, we want to compute projections, which are again partial functions.
Since our ultimate goal is to extract a correct implementation of EPP, we need to take a different approach to partiality and define
\begin{lstlisting}
Definition epp_C D ps C : projectable_C D C ps -> Network Sig'.
Definition epp_D D : projectable_D D -> DefSetB Sig'.
Definition epp P : projectable_P P -> Program Sig'.
\end{lstlisting}
taking a proof term as additional argument (for which we prove proof irrelevance).
These definitions are interactive, so we also state and prove lemmas showing that they yield the expected results as in pen-and-paper presentations~\cite{CM20,M22}.
\begin{lstlisting}
Lemma epp_C_Com_p : In p ps ->
  epp_C D ps (p # e --> qdollarx @ a;;C) HC p = q!e @!a; epp_C D ps C HC' p.
\end{lstlisting}

Paving the way for the EPP theorem, we prove a number of inversion lemmas for EPP, which cannot be trivially obtained by applying inversion to a hypothesis.
\begin{lstlisting}
Lemma epp_C_not_Branching_None_None : epp_C D ps C HC p <> q & None // None.

Lemma epp_C_Sel_Branching_l : epp_C D ps C HC p = q(+)left @+a Bp ->
  epp_C D ps C HC q = p & Bl // Br -> Bl <> None /\ Br = None.

Lemma epp_C_Cond_Send_inv : epp_C D ps (If p ?? b Then C1 Else C2) HC r = q!e @!a B ->
  exists B1 B2, epp_C D ps C1 HC1 r = q!e @!a B1
  /\ epp_C D ps C2 HC2 r = q!e @!a B2 /\ B1 [V] B2 == B.
\end{lstlisting}

\subsection{Strong projectability}
The operational correspondence between choreographies and their projections, which is the topic of \cref{sec:epp-thm}, states that a projectable choreography can make a transition iff its projection can make a corresponding transition.
Generalising this result to multi-step transitions requires chaining applications of this correspondence.
However, projectability is not preserved by transitions, due to how runtime terms are projected: \lstinline+RT_Call X ps C'+ is projected as \lstinline+Call (X,p)+ if \lstinline+p+ is in \lstinline+ps+, and as the projection of \lstinline+C'+ otherwise.
Our definition of projectability allows \lstinline+C'+ to be unprojectable for any process in \lstinline+ps+, which would make the result of the latter transition unprojectable.

This situation can never arise if one respects the intended usage of runtime terms: initially \lstinline+C'+ is the body of a procedure, and \lstinline+ps+ is the set of processes used in it.
Afterwards \lstinline+ps+ only shrinks, while \lstinline+C'+ may change due to execution of actions that involve processes not in \lstinline+ps+ (which keeps \coc{C'} projectable).
This assumption is implicit in pen-and-paper presentations.
We formalise it in the following definition of strong projectability.
\begin{lstlisting}
Fixpoint str_proj D (C:Choreography Sig) (r:Pid) : Prop :=
match C with
| eta @ a;; C' => str_proj D C' r
| If p ?? b Then C1 Else C2 => str_proj D C1 r /\ str_proj D C2 r /\ projectable_B D C r
| RT_Call X ps C => str_proj D C r /\ (forall p, In p ps ->
                      forall B B', [[D,snd (D X) | p]] == B -> [[D,C | p]] == B' -> B [>>] B')
| _ => True
end.
\end{lstlisting}
The last conjunct in the case of conditional is needed to guarantee that strong projectability implies projectability.
The last conjunct in the case of runtime terms captures the notion that \lstinline+C'+ may differ from the original definition of procedure \lstinline+X+, but the transitions in the reduction path did not involve processes that still have to execute the procedure call.

Projectability and strong projectability coincide for initial choreographies.
Furthermore, we state and prove lemmas that show that \lstinline+str_proj D C r+ implies both \lstinline+projectable_B D C r+ and \lstinline+str_proj D C' r+ for any choreography \lstinline+C'+ that \lstinline+C+ can transition to.
(This is the reason for including the last conjunct in the clause defining strong projectability of conditionals: without it, we still would not be able to prove that \lstinline+projectable_B D C' r+.)

Strong projectability for programs requires as expected that all choreographies in the program be strongly projectable.
Furthermore, we also require the program to be well-formed.
This assumption makes the definition simpler and more manageable, as all procedures will be initial and annotated with the right sets of processes.
\begin{lstlisting}
Definition str_proj_P P := Program_WF P /\ projectable_D (Procedures P) /\
  forall r, str_proj (Procedures P) (Main P) r.
\end{lstlisting}

Using these results, we can start relating the semantics of choreographies with the definition of EPP.
For example, if \lstinline+C+ can execute a communication from \lstinline+p+ to \lstinline+q+, then the behaviour of its projection for \lstinline+p+ starts by sending the corresponding expression to \lstinline+q+, while \lstinline+q+'s behaviour starts by receiving a value from \lstinline+p+.
\begin{lstlisting}
Lemma CCC_To_bproj_Com_p :
  str_proj D C p -> <<C,s>> --[RL_Com p v q x,D]--> <<C',s'>> ->
  exists e a Bp, [[D,C | p]] == Send Sig' q e a Bp /\ [[D,C' | p]] == Bp
              /\ v = eval_on_state Ev e s p.

Lemma CCC_To_bproj_Com_q :
  str_proj D C q -> <<C,s>> --[RL_Com p v q x,D]--> <<C',s'>> ->
  p <> q -> exists a Bq, [[D,C | q]] == Recv Sig' p x a Bq /\ [[D,C' | q]] == Bq.
\end{lstlisting}

An interesting corner case is what happens for processes not involved in the transition: they may lose some subbehaviours in branching terms due to some branches of conditionals disappearing from the choreography.
\begin{lstlisting}
Lemma CCC_To_bproj_disjoint :
  (forall X, CCC_pn (snd (D X)) (Names D) [C] fst (D X)) ->
  str_proj D C p -> disjoint_p_rl p tl -> <<C,s>> --[tl,D]--> <<C',s'>> ->
  exists B B', [[D,C | p]] == B /\ [[D,C' | p]] == B' /\ B [>>] B'.
\end{lstlisting}
The first hypothesis states that the procedures in \coc{D} are well-annotated.

As a consequence of these lemmas, we get that strong projectability is preserved by transitions.
\begin{lstlisting}
Lemma CCC_To_str_proj :
  (forall p, str_proj D C p) ->
  (forall p Y, str_proj D (snd (D Y)) p) ->
  (forall Y, CCC_pn (snd (D Y)) (Names D) [C] fst (D Y)) ->
  <<C,s>> --[t,D]--> <<C',s'>> -> forall p, str_proj D C' p.

Lemma CCP_To_str_proj : str_proj_P P -> (P,s) --[tl]--> (P',s') -> str_proj_P P'.
\end{lstlisting}
The hypotheses of the first lemma all hold if \coc{(D,C)} is a strongly projectable program.

\subsection{Discussion}
\label{sec:epp-discussion}

\paragraph{Modelling of partial functions.}
The definition of \lstinline+merge+ very explicitly considers the $2^4=16$ possible combinations of behaviours that can be offered when both arguments are branching terms.
This clearly does not scale if the set of labels is larger, and it is the place where our design choice of fixing it to a two-element set is most critical.
The same issue, but with a smaller impact, arises in the definition of \lstinline+more_branches+, which includes $2^2=4$ clauses related to branching terms, and in the definition of \lstinline+bproj+, which includes one clause for each branching term.

We do not think that this issue can be circumvented.
Our original approach considered an unspecified type \lstinline+Label:DecType+, and branching terms had type
\begin{lstlisting}
Branching : Pid -> (Label -> option (Ann*Behaviour)) -> Behaviour
\end{lstlisting}
This definition quickly proved unusable in practice: the induction principles generated by Coq were too weak, and most datatypes related to the process calculus had an undecidable equality.
Furthermore we ran into problems with definitions that required inspecting the behaviours associated to the labels because of the size restrictions in elimination combinators.

The first time we managed to have a working definition of \lstinline+merge+ was after fixing the set of labels to contain two elements.
This was the approach presented in~\cite{CMP21b}, where \lstinline+merge+ is formalised by first defining a total function
\begin{lstlisting}
Xmerge : XBehaviour -> XBehaviour -> XBehaviour
\end{lstlisting}
(where \lstinline+XBehaviour+ is a type including \lstinline+XUndefined+ subterms with the obvious intended meaning), and then defining \lstinline+merge B1 B2+ as \lstinline+Xmerge (inject B1) (inject B2)+, where \lstinline+inject+ is the trivial injection from \lstinline+Behaviour+ to \lstinline+XBehaviour+.

Apart from the added complexity of having duplications of types and definitions throughout the formalisation, working with these functions is very cumbersome.
The definition of \lstinline+Xmerge+ relies heavily on deciding equalities, so proofs of results about \lstinline+Xmerge+ necessarily had to perform the same eliminations.
At the end of the day, the number of cases in proofs was in the same order of magnitude as in the current version -- but they were generated in several verbose elimination steps, rather than directly from performing induction/inversion on a hypothesis.
Furthermore, the old definition required us to consider a significant number of absurd cases (in some lemmas, around $90\%$ of the total), whereas with the current definition these cases are simply not generated.
The only added complexity we noticed while adapting the formalisation was that we occasionally needed to apply lemma \lstinline+merge_unique+ to infer that two behaviours are identical -- but the size of this part of the formalisation was reduced by about $80\%$ (from around 3150 lines down to 700 lines).

Taking all these aspects into account, we believe that the current design choices are the most suitable for our theory.

\paragraph{Projectability.}
The lemmas relating projectability to the low-level semantics of choreographies typically include several hypotheses, cf.\ lemma \lstinline+CCC_To_str_proj+.
For programs, we packaged these properties in a single definition (\coc{str_proj_P}).
For the lower-level lemmas, we decided against this because not all these properties are needed in all lemmas -- some are only required in results involving procedure calls, others are important for conditionals, and communications require far fewer.
By including only the necessary assumptions in each lemma, we obtain more robust results.

\paragraph{Strong projectability.}
The need for strong projectability was independently identified in the pen-and-paper presentation in \cite{M22}.
There, the projectability requirement on runtime terms was included in the notion of well-formedness for choreographies.
While this option matches the intuition of ``intended usage of runtime terms'', it requires having defined projection.
In our formalisation, we strive for modularity, and we opted for a design where the choreographic calculus is fully decoupled from the target language and the definition of projection.
In this way, we allow for future extensions of our development with alternative definitions of EPP.

In the future, it would be interesting to investigate whether there is a syntactic characterisation of ``intended usage of runtime terms'' that is completely at the level of choreographies.
Such a characterisation would yield the benefits of both approaches described above: it would give us a notion of well-formedness closer to intuition, while keeping it decoupled from EPP.

\paragraph{Summary.}
The definitions of branching order, merge and Endpoint Projection, together with the accompanying lemmas, are divided in three files totaling 14 definitions, 126 lemmas and 15 tactics to automate recurring types of goals.
By far the largest bulk is the formalisation of EPP, at over 2200 lines of Coq code (with approximately 100 lemmas), while the branching order and merge require respectively 260 and 440 lines of Coq code (with a total of only 3 results that require longer proofs).

\section{The EPP theorem}
\label{sec:epp-thm}

The operational correspondence between choreographies and their projections, in languages that include both conditionals and out-of-order execution, is not as straightforward as for the simple language in \Cref{sec:background}.
In particular, branching terms in networks may linger for a bit longer compared to the choreographies that generated them.
This requires referring to the branching order in the EPP theorem:
\begin{lstlisting}
Lemma EPP_Complete : str_proj_P P -> (P,s) --[tl]--> (P',s') ->
  exists N tl', (epp P HP,s) --[tl']--> (N,s')
  /\ Procs N = Procs (epp P HP) /\ forall HP', Net N (>>) Net (epp P' HP').

Lemma EPP_Sound : str_proj_P P -> (epp P HP,s) --[tl]--> (N',s') ->
  exists P' tl', (P,s) --[tl']--> (P',s') /\ forall HP', Net N' (>>) Net (epp P' HP').
\end{lstlisting}
(Recall that \lstinline+epp+ takes a proof of projectability as its last argument.)

Completeness is not too hard to prove.
As in~\cite{CM20,M22}, the result is proven by considering the possible transitions that \lstinline+Main P+ can make; there are four cases, and the results proved earlier about the shape of the projection of \lstinline+P+ suffice to establish the thesis without too much work.
The whole proof is 250 lines long, and the generalisation to multi-step transitions requires an additional 40 lines.

The proof of soundness is known to be harder \cite{CHY12,M13p,CM20,M22}.
A common strategy is to proceed by induction on the choreography, and then do case analysis on the possible network transition.
The latter is either the first term in the choreography, and we can apply the matching choreography rule; or it is not, and we can apply a delay rule and invoke the induction hypothesis.

Each of these cases is challenging in itself, and they are therefore stated as separate lemmas on transitions.
As an example, the transition lemma for communications reads
\begin{lstlisting}
Lemma SP_To_bproj_Com : str_proj_P (D,C) ->
  <<epp_C D ps C HC,s>> --[RL_Com p v q x,D']--> <<N',s'>> ->
  exists C', <<C,s>> --[RL_Com p v q x,D]--> <<C',s'>> /\ forall HC', (N' (==) (epp_C D ps C' HC')).
\end{lstlisting}
and the corresponding proof script is around 320 lines long.
There are five of these lemmas in total, of a similar level of complexity.

Soundness also requires an additional lemma on procedure calls:
\begin{lstlisting}
Lemma SP_To_bproj_Call_name : <<epp_C D ps C HC,s>> --[RL_Call X p,D']--> <<N',s'>> ->
  exists (Y:RecVar), X = (Y,p) /\ X_Free _ Y C.
\end{lstlisting}
which is needed to apply the corresponding transition lemma.

Chaining applications of \lstinline+EPP_Sound+ also requires that extending the projection of a choreography with extra branches does not add transitions.
\begin{lstlisting}
Lemma SP_To_MBN_epp : N1 (>>) epp_C D' ps C HC -> <<N1,s>> --[tl,D]--> <<N2,s'>> ->
  exists N2', <<epp_C D' ps C HC,s>> --[tl,D]--> <<N2',s'>> /\ N2 (>>) N2'.
\end{lstlisting}

This result is lifted to \lstinline+SPP_To+ and \lstinline+SPP_ToStar+.
The latter generalisation requires applying \lstinline+EPP_Sound+.
It is then itself used to prove soundness of EPP for multi-step transitions.

The proof of the EPP theorem consists of an additional 2650 lines of Coq code, for only 14 lemmas.

\section{Related work}
\label{sec:rw}

The need for formalising concurrency theory is identified in \cite{MS15}, where the authors formalised a published article on a process calculus in Coq and discovered several major flaws in the proofs.
The authors
\begin{quote}
[\ldots] feel that it is [the errors'] very presence
in a peer-reviewed, state-of-the-art paper that strongly underlines the need for a more precise formal treatment of proofs in this domain. \cite[Section 6]{MS15}
\end{quote}
Since then, there have been a number of formalisation efforts in this area.
We discuss the ones closest to our work.

To the best of our knowledge, our original presentations \cite{CMP21a,CMP21b} were the first formalisations of a choreographic language featuring the expected programming constructs that allow for infinite and branching concurrent behaviour.
As we discussed, this article presents a substantial improvement of the original development.

More recently, there have been two additions to the family of fully-formalised choreographic programming languages.

Kalas is a certified compiler written in HOL from a choreographic language similar to ours to CakeML \cite{PGSN22}.
It includes an asynchronous semantics, but the notion of EPP is more restrictive than ours: it is an \emph{ad-hoc} definition that bypasses the need for the merge operator, but does not provide its full flexibility.
In particular, processes evaluating conditionals must immediately send selections to the processes that need them, while CC is more faithful to the pen-and-paper literature on choreographies \cite{CHY12,CM13,HYC16}.

Pirouette is a functional choreographic programming language formalised in Coq \cite{HG22}.
It supports asynchronous communication and higher-order functions, but at the cost of introducing hidden global synchronisations for \emph{all} processes whenever a function is called.
The semantics of CC is, instead, decentralised and all synchronisations are syntactically explicit.

Extending CC with asynchronous communication has also been studied \cite{CM17ice}, but since it was not part of the reference pen-and-paper work that we followed, we postponed its formalisation to future work.

Another line of research connected to choreographic programming is that of multiparty session types \cite{HYC16}.
These types are essentially choreographies without computation (e.g., communications only specify sender, receiver, and message type, but not how the message is computed or where it is stored), and are therefore simpler than CC.
There are two available formalisations of multiparty session types \cite{CFGY21,JBK22}.
Both formalisations include a counterpart to the EPP theorem, but they are even more restrictive than Kalas in how they handle the projection of conditionals.

\section{Conclusion}
\label{sec:discussion}
\label{sec:concl}

We presented a formalisation of a state-of-the-art article on theory of choreographic programming.
The formalisation process unveiled subtle problems in definitions, making a case for a more systematic use of theorem provers to validate results in the field.
Even more, it positively impacted the theory itself, showing that formalisation can be valuable tool also in the design phase of the research process.

Our formalisation was done in parallel with the pen-and-paper revision of CC carried out in \cite{M22}.
There are two interesting observations to make about this parallel development.
First, many of the technical aspects that we discuss in this article were also independently discovered during the writing of \cite{M22}.
Second, the seemingly disparate goals of making the theory more intuitive to students and amenable to formalisation actually converged on the same solution, and sometimes resulted in useful exchanges of feedback.
Taken together, these two observations strongly suggest that the current formulation of CC is the ``right'' one, and offers a suitable basis for future developments.

We have already started exploring extensions and applications of our formalisation.
These include amendment (a procedure that injects appropriate selections to make a choreography projectable), a proof of starvation-freedom, alternative definitions of EPP, and applying program extraction to develop a certified toolchain from choreographies to executable code.

An important tool for future extensions is stronger automation for proofs about choreographies.
Our development already includes a few simple tactics that deal with commonly-recurring goals, but it would be worthwhile to extend this library with more powerful tactics, e.g., to reason about multi-step transitions.
Furthermore, for many proofs by structural induction, there are strong similarities among their different cases, and it would be interesting to try to automate proof strategies that can capitalise on this.

The appeal of choreographic programming largely depends on its promise of delivering correct implementations, by removing the possibility of human error through EPP.
This promise has motivated a proliferation of choreographic programming languages, including features of practical value such as asynchronous communication, nondeterminism, broadcast, dynamic network topologies, and more~\cite{Aetal16,GVWY17,Hetal16,M22}.
The theories of these languages are becoming more and more complex, thus increasing the likelihood of critical mistakes and making the case for more trustworthy developments.
We hope that our work can contribute a solid foundation for the development of these features.

\begin{acknowledgements}
  This work was partially supported by Villum Fonden, grant no. 29518, and the Danish Council for Independent Research, Natural
  Sciences, grant no. 0135-00219.
\end{acknowledgements}

\bibliographystyle{spmpsci}
\bibliography{biblio}

\begin{thebibliography}{10}
\providecommand{\url}[1]{{#1}}
\providecommand{\urlprefix}{URL }
\expandafter\ifx\csname urlstyle\endcsname\relax
  \providecommand{\doi}[1]{DOI~\discretionary{}{}{}#1}\else
  \providecommand{\doi}{DOI~\discretionary{}{}{}\begingroup
  \urlstyle{rm}\Url}\fi

\bibitem{forte2016}
Albert, E., Lanese, I. (eds.): Formal Techniques for Distributed Objects,
  Components, and Systems - 36th {IFIP} {WG} 6.1 International Conference,
  {FORTE} 2016, Held as Part of the 11th International Federated Conference on
  Distributed Computing Techniques, DisCoTec 2016, Heraklion, Crete, Greece,
  June 6-9, 2016, Proceedings, \emph{Lecture Notes in Computer Science}, vol.
  9688. Springer (2016)

\bibitem{Aetal16}
Ancona, D., Bono, V., Bravetti, M., Campos, J., Castagna, G., Deni{\'{e}}lou,
  P., Gay, S.J., Gesbert, N., Giachino, E., Hu, R., Johnsen, E.B., Martins, F.,
  Mascardi, V., Montesi, F., Neykova, R., Ng, N., Padovani, L., Vasconcelos,
  V.T., Yoshida, N.: Behavioral types in programming languages.
\newblock Foundations and Trends in Programming Languages \textbf{3}(2--3),
  95--230 (2016)

\bibitem{CP10}
Caires, L., Pfenning, F.: Session types as intuitionistic linear propositions.
\newblock In: P.~Gastin, F.~Laroussinie (eds.) Procs.\ CONCUR, \emph{Lecture
  Notes in Computer Science}, vol. 6269, pp. 222--236. Springer (2010).
\newblock \doi{10.1007/978-3-642-15375-4\_16}

\bibitem{CHY12}
Carbone, M., Honda, K., Yoshida, N.: Structured communication-centered
  programming for web services.
\newblock {ACM} Trans.\ Program.\ Lang.\ Syst. \textbf{34}(2), 8:1--8:78
  (2012).
\newblock \doi{10.1145/2220365.2220367}

\bibitem{CM13}
Carbone, M., Montesi, F.: Deadlock-freedom-by-design: multiparty asynchronous
  global programming.
\newblock In: R.~Giacobazzi, R.~Cousot (eds.) Procs.\ POPL, pp. 263--274. {ACM}
  (2013).
\newblock \doi{10.1145/2429069.2429101}

\bibitem{CDP11}
Castagna, G., Dezani{-}Ciancaglini, M., Padovani, L.: On global types and
  multi-party sessions.
\newblock In: R.~Bruni, J.~Dingel (eds.) Procs.\ FORTE, \emph{LNCS}, vol. 6722,
  pp. 1--28. Springer (2011).
\newblock \doi{10.1007/978-3-642-21461-5\_1}

\bibitem{CFGY21}
Castro{-}Perez, D., Ferreira, F., Gheri, L., Yoshida, N.: Zooid: a {DSL} for
  certified multiparty computation: from mechanised metatheory to certified
  multiparty processes.
\newblock In: S.N. Freund, E.~Yahav (eds.) Procs.\ PLDI, pp. 237--251. {ACM}
  (2021).
\newblock \doi{10.1145/3453483.3454041}

\bibitem{CLM17}
Cruz{-}Filipe, L., Larsen, K.S., Montesi, F.: The paths to choreography
  extraction.
\newblock In: J.~Esparza, A.S. Murawski (eds.) Procs.\ FOSSACS, \emph{LNCS},
  vol. 10203, pp. 424--440 (2017).
\newblock \doi{10.1007/978-3-662-54458-7\_25}

\bibitem{CM16}
Cruz{-}Filipe, L., Montesi, F.: Choreographies in practice.
\newblock In: Albert and Lanese  \cite{forte2016}, pp. 114--123.
\newblock \doi{10.1007/978-3-319-39570-8\_8}

\bibitem{CM17ice}
Cruz{-}Filipe, L., Montesi, F.: On asynchrony and choreographies.
\newblock In: M.~Bartoletti, L.~Bocchi, L.~Henrio, S.~Knight (eds.) Procs.\
  ICE, \emph{{EPTCS}}, vol. 261, pp. 76--90 (2017).
\newblock \doi{10.4204/EPTCS.261.8}

\bibitem{CM17}
Cruz{-}Filipe, L., Montesi, F.: Procedural choreographic programming.
\newblock In: A.~Bouajjani, A.~Silva (eds.) Procs.\ FORTE, \emph{Lecture Notes
  in Computer Science}, vol. 10321, pp. 92--107. Springer (2017).
\newblock \doi{10.1007/978-3-319-60225-7\_7}

\bibitem{CM20}
Cruz{-}Filipe, L., Montesi, F.: A core model for choreographic programming.
\newblock Theor.\ Comput.\ Sci. \textbf{802}, 38--66 (2020).
\newblock \doi{10.1016/j.tcs.2019.07.005}

\bibitem{CMP19}
Cruz-Filipe, L., Montesi, F., Peressotti, M.: Choreographies in {Coq}.
\newblock In: TYPES 2019, Abstracts (2019).
\newblock Extended abstract

\bibitem{CMP21b}
Cruz{-}Filipe, L., Montesi, F., Peressotti, M.: Certifying choreography
  compilation.
\newblock In: A.~Cerone, P.C. {\"{O}}lveczky (eds.) Procs.\ ICTAC, \emph{LNCS},
  vol. 12819, pp. 115--133. Springer (2021).
\newblock \doi{10.1007/978-3-030-85315-0\_8}

\bibitem{CMP21a}
Cruz{-}Filipe, L., Montesi, F., Peressotti, M.: Formalising a {Turing-}complete
  choreographic language in {Coq}.
\newblock In: L.~Cohen, C.~Kaliszyk (eds.) Procs.\ ITP, \emph{LIPIcs}, vol.
  193, pp. 15:1--15:18. Schloss Dagstuhl -- Leibniz-Zentrum f{\"{u}}r
  Informatik (2021).
\newblock \doi{10.4230/LIPIcs.ITP.2021.15}

\bibitem{CMP22-source}
Cruz-Filipe, L., Montesi, F., Peressotti, M.: A formal theory of choreographic
  programming in {Coq} (2022).
\newblock \doi{10.5281/zenodo.7050062}.
\newblock \urlprefix\url{https://doi.org/10.5281/zenodo.7050062}

\bibitem{DGGLM17}
Dalla~Preda, M., Gabbrielli, M., Giallorenzo, S., Lanese, I., Mauro, J.:
  Dynamic choreographies: Theory and implementation.
\newblock Log.\ Methods Comput.\ Sci. \textbf{13}(2) (2017).
\newblock \doi{10.23638/LMCS-13(2:1)2017}

\bibitem{GVWY17}
Gay, S.J., Vasconcelos, V.T., Wadler, P., Yoshida, N.: Theory and applications
  of behavioural types (dagstuhl seminar 17051).
\newblock Dagstuhl Reports \textbf{7}(1), 158--189 (2017).
\newblock \doi{10.4230/DagRep.7.1.158}

\bibitem{GLR18}
Giallorenzo, S., Lanese, I., Russo, D.: Chip: {A} choreographic integration
  process.
\newblock In: H.~Panetto, C.~Debruyne, H.A. Proper, C.A. Ardagna, D.~Roman,
  R.~Meersman (eds.) Procs.\ OTM, part II, \emph{Lecture Notes in Computer
  Science}, vol. 11230, pp. 22--40. Springer (2018).
\newblock \doi{10.1007/978-3-030-02671-4\_2}

\bibitem{GMP20}
Giallorenzo, S., Montesi, F., Peressotti, M.: Choreographies as objects.
\newblock CoRR \textbf{abs/2005.09520} (2020).
\newblock \urlprefix\url{https://arxiv.org/abs/2005.09520}

\bibitem{bpmn}
{O}bject~{M}anagement {G}roup: {B}usiness {P}rocess {M}odel and {N}otation.
\newblock
  \href{http://www.omg.org/spec/BPMN/2.0/}{http://www.omg.org/spec/BPMN/2.0/}
  (2011)

\bibitem{HG22}
Hirsch, A.K., Garg, D.: Pirouette: higher-order typed functional
  choreographies.
\newblock Proc.\ {ACM} Program.\ Lang. \textbf{6}({POPL}), 1--27 (2022).
\newblock \doi{10.1145/3498684}

\bibitem{HYC16}
Honda, K., Yoshida, N., Carbone, M.: Multiparty asynchronous session types.
\newblock J. {ACM} \textbf{63}(1), 9 (2016).
\newblock \doi{10.1145/2827695}.
\newblock Also: POPL, pages 273--284, 2008

\bibitem{Hetal16}
H{\"{u}}ttel, H., Lanese, I., Vasconcelos, V.T., Caires, L., Carbone, M.,
  Deni{\'{e}}lou, P., Mostrous, D., Padovani, L., Ravara, A., Tuosto, E.,
  Vieira, H.T., Zavattaro, G.: Foundations of session types and behavioural
  contracts.
\newblock {ACM} Comput.\ Surv. \textbf{49}(1), 3:1--3:36 (2016).
\newblock \doi{10.1145/2873052}

\bibitem{msc}
{Intl. Telecommunication Union}: Recommendation \mbox{Z.120}: {Message Sequence
  Chart} (1996)

\bibitem{JBK22}
Jacobs, J., Balzer, S., Krebbers, R.: Multiparty gv: Functional multiparty
  session types with certified deadlock freedom.
\newblock In: Procs.\ ICFP (2022).
\newblock Accepted for publication

\bibitem{Kleene52}
Kleene, S.C.: Introduction to Metamathematics, vol.~1.
\newblock North-Holland Publishing Co. (1952)

\bibitem{LN15}
Lluch{-}Lafuente, A., Nielson, F., Nielson, H.R.: Discretionary information
  flow control for interaction-oriented specifications.
\newblock In: N.~Mart{\'{\i}}{-}Oliet, P.C. {\"{O}}lveczky, C.L. Talcott (eds.)
  Logic, Rewriting, and Concurrency, \emph{Lecture Notes in Computer Science},
  vol. 9200, pp. 427--450. Springer (2015).
\newblock \doi{10.1007/978-3-319-23165-5\_20}

\bibitem{LH17}
L{\'{o}}pez, H.A., Heussen, K.: Choreographing cyber-physical distributed
  control systems for the energy sector.
\newblock In: A.~Seffah, B.~Penzenstadler, C.~Alves, X.~Peng (eds.) Procs.\
  SAC, pp. 437--443. {ACM} (2017).
\newblock \doi{10.1145/3019612.3019656}

\bibitem{LNN16}
L{\'{o}}pez, H.A., Nielson, F., Nielson, H.R.: Enforcing availability in
  failure-aware communicating systems.
\newblock In: Albert and Lanese  \cite{forte2016}, pp. 195--211.
\newblock \doi{10.1007/978-3-319-39570-8\_13}

\bibitem{MS15}
Maksimovic, P., Schmitt, A.: {HOCore in Coq}.
\newblock In: C.~Urban, X.~Zhang (eds.) Procs.\ ITP, \emph{Lecture Notes in
  Computer Science}, vol. 9236, pp. 278--293. Springer (2015).
\newblock \doi{10.1007/978-3-319-22102-1\_19}

\bibitem{M13p}
Montesi, F.: Choreographic programming.
\newblock {Ph.{D}. Thesis}, IT University of Copenhagen (2013).
\newblock
  \urlprefix\url{http://www.fabriziomontesi.com/files/choreographic\_programming.pdf}

\bibitem{M22}
Montesi, F.: Introduction to choreographies (2022).
\newblock Accepted for publication by Cambridge University Press

\bibitem{NS78}
Needham, R.M., Schroeder, M.D.: Using encryption for authentication in large
  networks of computers.
\newblock Commun.\ {ACM} \textbf{21}(12), 993--999 (1978).
\newblock \doi{10.1145/359657.359659}

\bibitem{PGSN22}
Pohjola, J.{\AA}., G{\'{o}}mez{-}Londo{\~{n}}o, A., Shaker, J., Norrish, M.:
  Kalas: {A} verified, end-to-end compiler for a choreographic language.
\newblock In: J.~Andronick, L.~de~Moura (eds.) Procs.\ ITP, \emph{LIPIcs}, vol.
  237, pp. 27:1--27:18. Schloss Dagstuhl - Leibniz-Zentrum f{\"{u}}r Informatik
  (2022).
\newblock \doi{10.4230/LIPIcs.ITP.2022.27}

\bibitem{SY19}
Scalas, A., Yoshida, N.: Less is more: multiparty session types revisited.
\newblock Proc.\ {ACM} Program.\ Lang. \textbf{3}({POPL}), 30:1--30:29 (2019).
\newblock \doi{10.1145/3290343}

\bibitem{wscdl}
{W3C}: {WS Choreography Description Language}.
\newblock
  \href{http://www.w3.org/TR/ws-cdl-10/}{http://www.w3.org/TR/ws-cdl-10/}
  (2004)

\end{thebibliography}

\end{document}